\documentclass[%
twocolumn,
floatfix,
amsmath,amssymb,
aps,
superscriptaddress,
pra,
]{revtex4-2}

\usepackage[svgnames,psnames]{xcolor}
\usepackage{url}
\usepackage{graphicx}
\graphicspath{{.}{figs/}{../figs/}}
\usepackage{afterpage}
\usepackage{dcolumn}
\usepackage{bm}
\usepackage{tikz}
\usetikzlibrary{positioning,intersections,angles,quotes,quantikz}
\usepackage[caption=false]{subfig}
\usepackage{booktabs} 
\usepackage{comment}
\usepackage{physics}
\usepackage{mathtools}

\usepackage[colorlinks,citecolor=DarkGreen,linkcolor=FireBrick,linktocpage,unicode,hypertexnames=false]{hyperref}
\usepackage[capitalize]{cleveref}
\crefname{enumi}{}{}

\begin{document}
\title{Simulation and performance analysis of quantum error correction with a rotated surface code under a realistic noise model
}
\author{Mitsuki Katsuda}
\email{u037351e@alumni.osaka-u.ac.jp}
\affiliation{%
  Graduate School of Engineering Science, Osaka University, 1-3 Machikaneyama, Toyonaka, Osaka 560-8531, Japan
}%
\author{Kosuke Mitarai}%
\email{mitarai@qc.ee.es.osaka-u.ac.jp}
\affiliation{%
  Graduate School of Engineering Science, Osaka University, 1-3 Machikaneyama, Toyonaka, Osaka 560-8531, Japan
}%
\affiliation{%
  Center for Quantum Information and Quantum Biology,
  Osaka University, 1-2 Machikaneyama, Toyonaka 560-0043, Japan
}%
\affiliation{%
  JST,
  PRESTO,
  4-1-8 Honcho, Kawaguchi, Saitama 332-0012, Japan
}%
\author{Keisuke Fujii}%
\email{fujii@qc.ee.es.osaka-u.ac.jp}
\affiliation{%
   Graduate School of Engineering Science, Osaka University, 1-3 Machikaneyama, Toyonaka, Osaka 560-8531, Japan
}%
\affiliation{%
  Center for Quantum Information and Quantum Biology,
  Osaka University, 1-2 Machikaneyama, Toyonaka 560-0043, Japan
}%
\affiliation{%
  RIKEN Center for Quantum Computing (RQC),
  Hirosawa 2-1, Wako, Saitama 351-0198, Japan
}%
\affiliation{%
  Fujitsu Quantum Computing Joint Research Division at QIQB,
  Osaka University, 1-2 Machikaneyama, Toyonaka 560-0043, Japan
}%

\date{\today}
\begin{abstract}
The demonstration of quantum error correction (QEC) is one of the most important milestones in the realization of fully-fledged quantum computers. Toward this, QEC experiments using the surface codes have recently been actively conducted. However, it has not yet been realized to protect logical quantum information beyond the physical coherence time. In this work, we performed a full simulation of QEC for the rotated surface codes with a code distance 5, which employs 49 qubits and is within reach of the current state-of-the-art quantum computers.
In particular, we evaluate the logical error probability in a realistic noise model that incorporates not only stochastic Pauli errors but also coherent errors due to a systematic control error or unintended interactions.
While a straightforward simulation of 49 qubits is not tractable within a reasonable computational time, 
we reduced the number of qubits required to 26 qubits by delaying the syndrome measurement in simulation.
This and a fast quantum computer simulator, Qulacs, implemented on GPU
allows us to simulate full QEC with an arbitrary local noise within reasonable simulation time.
Based on the numerical results, 
we also construct and verify an effective model to incorporate the effect of the coherent error into a stochastic noise model. This allows us to understand what the effect coherent error has on the logical error probability on a large scale without full simulation
based on the detailed full simulation of a small scale.
The present simulation framework and effective model, which can handle arbitrary local noise, will play a vital role in clarifying the physical parameters that future experimental QEC should target.
\end{abstract}
\maketitle
\section{Introduction}
        Quantum error correction (QEC) is essential for the realization of quantum computers
        because physical qubits suffer from errors due to decoherence caused by undesirable interactions with the environment.
        QEC protects them from such errors by encoding logical information of qubits on many physical qubits~\cite{shor1996fault}.
        With the progress of quantum hardware, it is becoming possible to precisely control tens to a hundred qubits, leading to experimental demonstrations of simple QEC codes on a variety of physical systems~\cite{zhang2012experimental,abobeih2021fault,egan2021fault,luo2021quantum,andersen2020repeated, ai2021exponential}.
        Surface codes proposed by Kitaev~\cite{dennis2002topological} has attracted much attention as a method to implement an error correction with superconducting qubits due to their relatively high threshold for local errors and their implementability using a two-dimensional lattice of qubits.
        By using the rotated surface code~\cite{horsman2012surface}, we can make one logical qubit with code distance $d$ using only $2d^2-1$ qubits.
        This means that we can implement $d=3$ and $d=5$ surface codes with 17 and 49 physical qubits, respectively.
        Currently available devices such as the ones presented in \cite{aruteQuantumSupremacyUsing2019, Wu2021} can handle qubits of this scale. Experimental efforts are underway to demonstrate  QEC with the surface codes using the superconducting qubits. 
        Ref.~\cite{ai2021exponential} has achieved exponential suppression of the logical error probability with the one-dimensional repetitive code,
        which is a one-dimensional substructure of the surface code.
        Moreover, Refs.~\cite{krinner2022realizing, zhao2022realization, Bluvstein2021} have implemented all of the operations necessary for the implementation of QEC in a two-dimensional surface code with $d=3$, while they do not achieve the break-even point, that is, the logical error probability is smaller than the physical error probability.
    
        To demonstrate that QEC can actually achieve a logical error probability lower than the physical one, it is essential to implement a surface code with a code distance of $d=5$ that can correct up to two errors.
        This is because two-qubit operations inevitably introduce two-qubit correlated errors on physical qubits.
        However, successful experimental realizations of surface codes \cite{corcoles2015demonstration, ai2021exponential, krinner2022realizing, zhao2022realization,Bluvstein2021} are still up to $d=3$~\cite{krinner2022realizing,zhao2022realization,Bluvstein2021} possibly due to the limited fidelity of gates and readouts in present quantum devices.
        To clarify what ultimately limits the experimental implementation of the error correction, detailed numerical analyses are required; we need to determine how much performance is necessary for which parameters, and how much the logical error probability can be reduced if they are achieved.

        Numerical experiments of surface codes have been carried out under various noise models.
        The simplest, albeit not truly realistic, model is the stochastic Pauli errors where Pauli operators act on the qubits probabilistically.
        In this case, the Gottesman-Knill theorem~\cite{nielsen2010quantum} allows us to simulate even large systems efficiently.
        In real experimental systems, unfortunately, there is a noise that has coherence and cannot be described by stochastic Pauli errors originating from, for example, over-rotation with a systematic control error, global external fields, cross-talk and so on. 
        
        Simulating QEC with these noises is as difficult as simulating a universal quantum computer in general. However, in certain limited cases, it is possible to numerically evaluate their performance.
        For example, 
        a mixture of coherent and incoherent noise on the one-dimensional repetition code has been analysed in detail by making use of its exact solvability mapping it to free-fermionic dynamics~\cite{suzuki2017efficient}.
        The $d=3$ rotated surface code with amplitude damping and dephasing~\cite{tomita2014low} has been analysed by exact simulation of the system, which is possible due to its small number of qubits.
        A sophisticated technique for contraction of 2D tensor networks has been used to simulate surface codes with arbitrary local noise on the data qubits, while the syndrome measurements are assumed to be ideal~\cite{darmawan2017tensor}.
        It is not clear if this method can be extended to simulate circuit-level noise, where each elementary operation is subject to noise.
        More recently, quasiprobability decomposition of non-Clifford channels into Clifford channels is utilized to simulate surface codes with a small coherent noise~\cite{hakkaku2021sampling}.
        This method is, however, not applicable to arbitrary noise because the coherence of the noise increases its sampling overhead exponentially.
        Given the finite precision of control systems in actual experiments, 
        the existence of coherent errors is inevitable.
        A detailed simulation of the surface code with $d=5$, which is the near-term milestone of QEC, is still challenging for classical computers. A framework to realize such simulations and to obtain knowledge on the impact of a coherent error on QEC is highly demanded.

        In this work, we fully simulate the QEC under a realistic noise model, including incoherent and coherent noise, in the $d=5$ rotated surface code with 49 physical qubits, and analyzed the effect of coherent errors on the logical error probability.
        The main obstacle to its analysis is that a straightforward simulation of 49 qubits would require a complex vector of dimension $2^{49}$.
        This prevents us from simulating the dynamics with a realistic computational resource.
        We overcome this obstacle by exploiting the structure of the syndrome measurement and reusing the measured qubits in the simulation.
        This allows us to achieve a full simulation of the $d=5$ rotated surface code by simulating only 26 qubits, thus making it feasible to analyze the effects of arbitrary local noise models on this QEC code.
        In particular, assuming its implementation on superconducting qubits, we use a realistic gate set and noise model, such as coherent errors in one-qubit operations and cross resonance gates in addition to naive stochastic Pauli errors.
        Moreover, we develop an effective model of physical error probability for incorporating the effects of coherent errors in rotated surface codes combining the simulation results and the previous analysis of coherent errors in 1D repetitive codes \cite{suzuki2017efficient}.
        Using this model, we investigate the possible regime of coherence time, gate time, coherent error ratio, etc., required for maintaining the quantum information of a logical qubit beyond the coherence time of a physical qubit.
        The results show that if the ratio of gate operation time to coherence time is below 0.005, the lifetime of the logical qubit exceeds that of the physical qubit, even if coherent errors occur with the same magnitude as their incoherent counterparts.
        On the other hand, it was also found that if the magnitude of the coherent error can be reduced to 20\% of the incoherent one, the ratio of gate operation time to coherence time is acceptable up to 0.007.
        The present simulation framework for QEC would provide an important guideline for future experimental demonstrations of QEC to extend the lifetime of logical qubits.

\section{Simulation methods for $d=5$ rotated surface code}\label{sec:simulation-method}
    \subsection{Circuits for syndrome measurements}
        \begin{figure}
            \centering
            \subfloat[\label{subfig:rotated-surface-code}]
            {
                \includegraphics[width=0.7\linewidth]{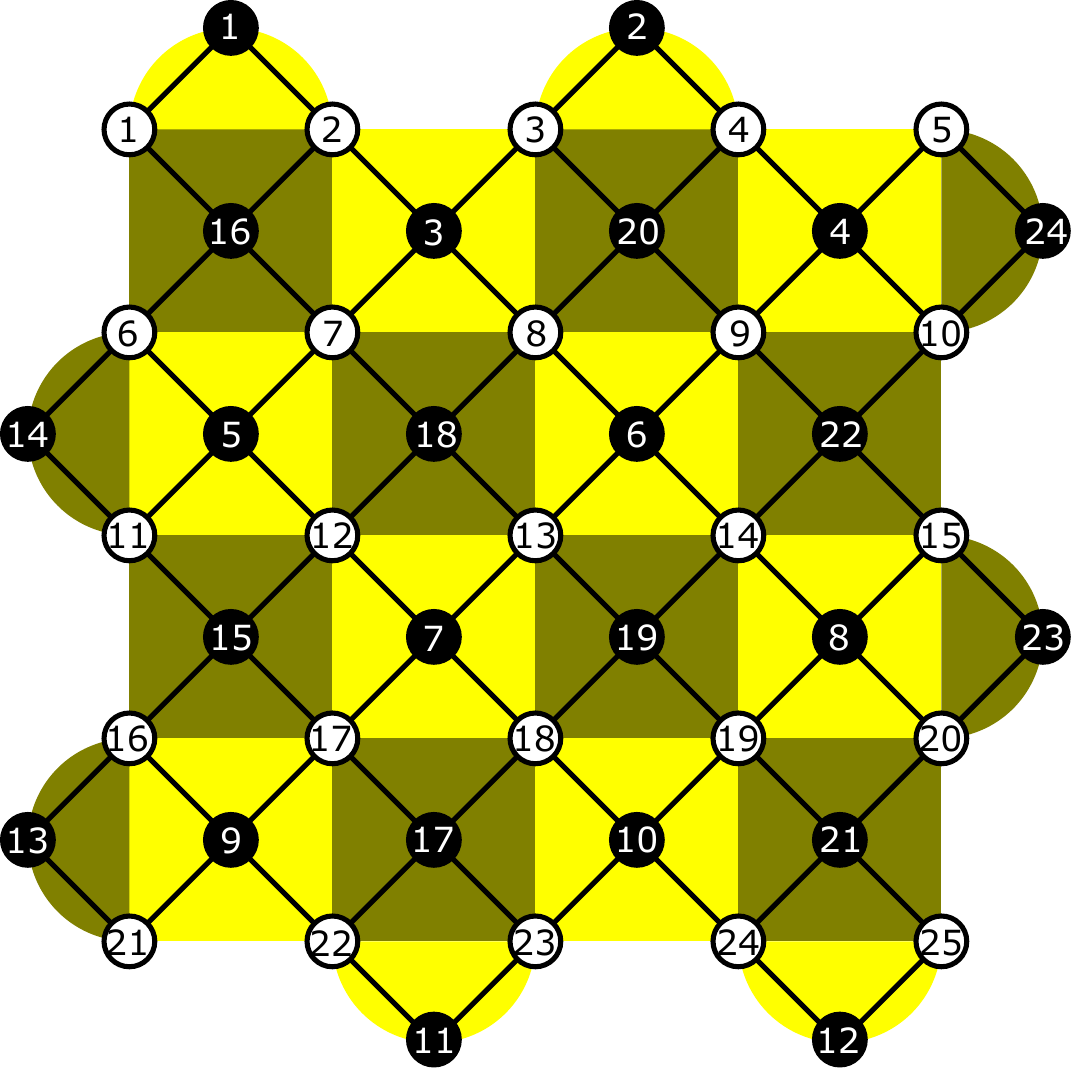}
            }\\
            \subfloat[\label{subfig:X-stabilizer}]
            {
                \includegraphics[scale=1.0]{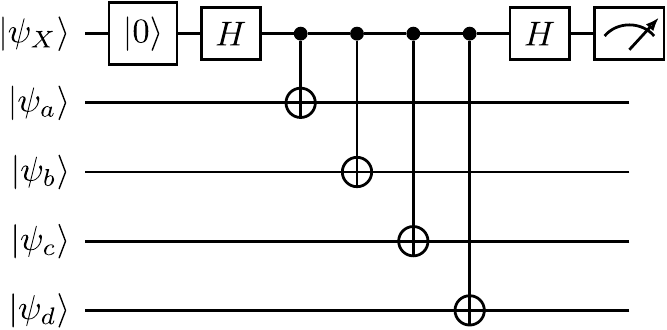}
            }\\
            \subfloat[\label{subfig:Z-stabilizer}]
            {
                \includegraphics[scale=1.0]{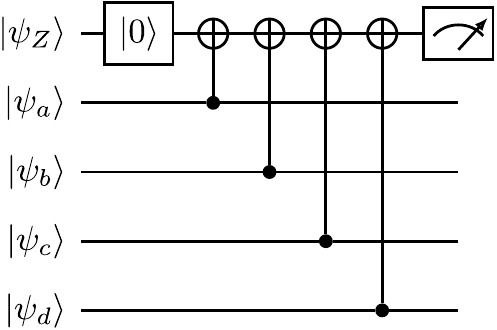}
            }
            \caption{\protect\subref{subfig:rotated-surface-code}The $d=5$ rotated surface codes. Yellow faces(\begin{tikzpicture}[baseline=0pt]\fill[yellow] (0, 0) rectangle (1em, 1em);\end{tikzpicture}) corresponds to $X$-type stabilizer and ocher ones(\begin{tikzpicture}[baseline=0pt]\fill[olive] (0, 0) rectangle (1em, 1em);\end{tikzpicture}) corresponds to $Z$-type stabilizer.White circles(\begin{tikzpicture}[baseline=0pt]\protect\draw (0, .4em) circle [radius=.5em];\end{tikzpicture}) means data qubits with their index, and black circles(\begin{tikzpicture}[baseline=0pt]\fill (0, .4em) circle [radius=.5em];\end{tikzpicture}) means ancilla qubits used for measurement and the order in which the measurement circuits are run. \protect\subref{subfig:X-stabilizer}Circuit for the $X$-type stabilizer measurement \protect\subref{subfig:Z-stabilizer}Circuit for the $Z$-type stabilizer measurement.}
            \label{fig:syndrome-measurement}
        \end{figure}
        
        \begin{table}[b]
            \caption{List of stabilizers of the $d=5$ rotated surface code}
            \begin{ruledtabular}
                \begin{tabular}{rcc}
                    Index & X-stabilizer & Z-stabilizer\\
                    \colrule
                    1 &  $X_{1}X_{2}$               & $Z_{16}Z_{21}$ \\
                    2 &  $X_{3}X_{4}$               & $Z_{6}Z_{11}$ \\
                    3 &  $X_{2}X_{3}X_{7}X_{8}$     & $Z_{11}Z_{12}Z_{16}Z_{17}$ \\
                    4 &  $X_{4}X_{5}X_{9}X_{10}$    & $Z_{1}Z_{2}Z_{6}Z_{7}$ \\
                    5 &  $X_{6}X_{7}X_{11}X_{12}$   & $Z_{17}Z_{18}Z_{22}Z_{23}$ \\
                    6 &  $X_{8}X_{9}X_{13}X_{14}$   & $Z_{7}Z_{8}Z_{12}Z_{13}$ \\
                    7 &  $X_{12}X_{13}X_{17}X_{18}$ & $Z_{13}Z_{14}Z_{18}Z_{19}$ \\
                    8 &  $X_{14}X_{15}X_{19}X_{20}$ & $Z_{3}Z_{4}Z_{8}Z_{9}$ \\
                    9 &  $X_{16}X_{17}X_{21}X_{22}$ & $Z_{19}Z_{20}Z_{24}Z_{25}$ \\
                    10 & $X_{18}X_{19}X_{23}X_{24}$ & $Z_{9}Z_{10}Z_{14}Z_{15}$ \\
                    11 & $X_{22}X_{23}$             & $Z_{15}Z_{20}$ \\
                    12 & $X_{24}X_{25}$             & $Z_{5}Z_{10}$ \\
                \end{tabular}
            \end{ruledtabular}
            \label{tab:stabilizers}
        \end{table}
        The $d=5$ rotated surface code,
        which is the target in this work, is shown in FIG.\ref{subfig:rotated-surface-code}.
        White circles and black circles in Fig.~\ref{subfig:rotated-surface-code} represent data qubits and ancilla qubits for the sydrome measurements, respectively.
        The list of stabilizers of this code is summarized in Table ~\ref{tab:stabilizers}.
        In this table, $X_i,\, Z_i$ mean Pauli operators acting on the $i$-th data qubit.
        These stabilizers are also illustrated in Fig.~\ref{subfig:rotated-surface-code}.
        $X$-type stabilizers correspond to yellow plaquettes, and $Z$-type stabilizers do to ocher ones.
        These stabilizers are measured with the circuits shown in Figs.~\ref{subfig:X-stabilizer} and ~\ref{subfig:Z-stabilizer}, respectively.
        
        The order of CNOT gates is important to determine the minimum depth of the syndrome measurement circuit.
        In this work, we choose the following order.
        In the case of $X$-type stabilizers, they are applied clockwise from the bottom right, while in the case of $Z$-type it is clockwise from the top right.
        For example, if $Z_1Z_2Z_6Z_7$ is to be measured, we apply CNOT gates with the order of 2,7,6,1.
        We have to make a special treatment for the measurement qubits on the boundary as they have fewer CNOT gates than the other measurement qubits.
        In this work, we make them ``wait'' if their target data qubit doesn't exist.
        While ``waiting``, one-qubit noise is applied.
        For example, the sequence for measuring $X_1X_2$ is ``2, 1, wait, wait'', and for measuring $X_{22}X_{23}$ is ``wait, wait, 22, 23''.

    \subsection{Reducing the number of qubits for simulation}
        \begin{figure}
            \subfloat[\label{subfig:repetition-1}]
            {
                \includegraphics[scale=0.4]{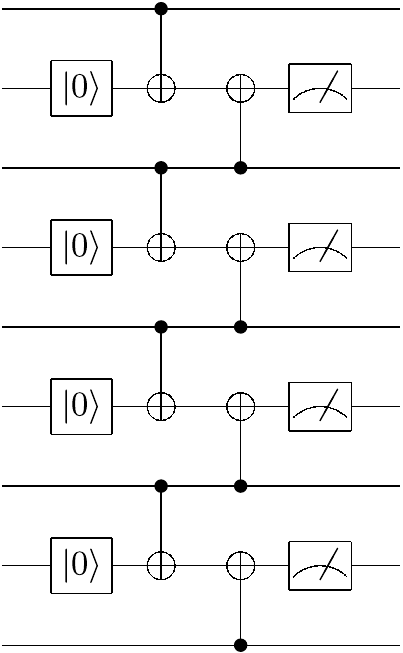}
            }
            \subfloat[\label{subfig:repetition-2}]
            {
                \includegraphics[scale=0.4]{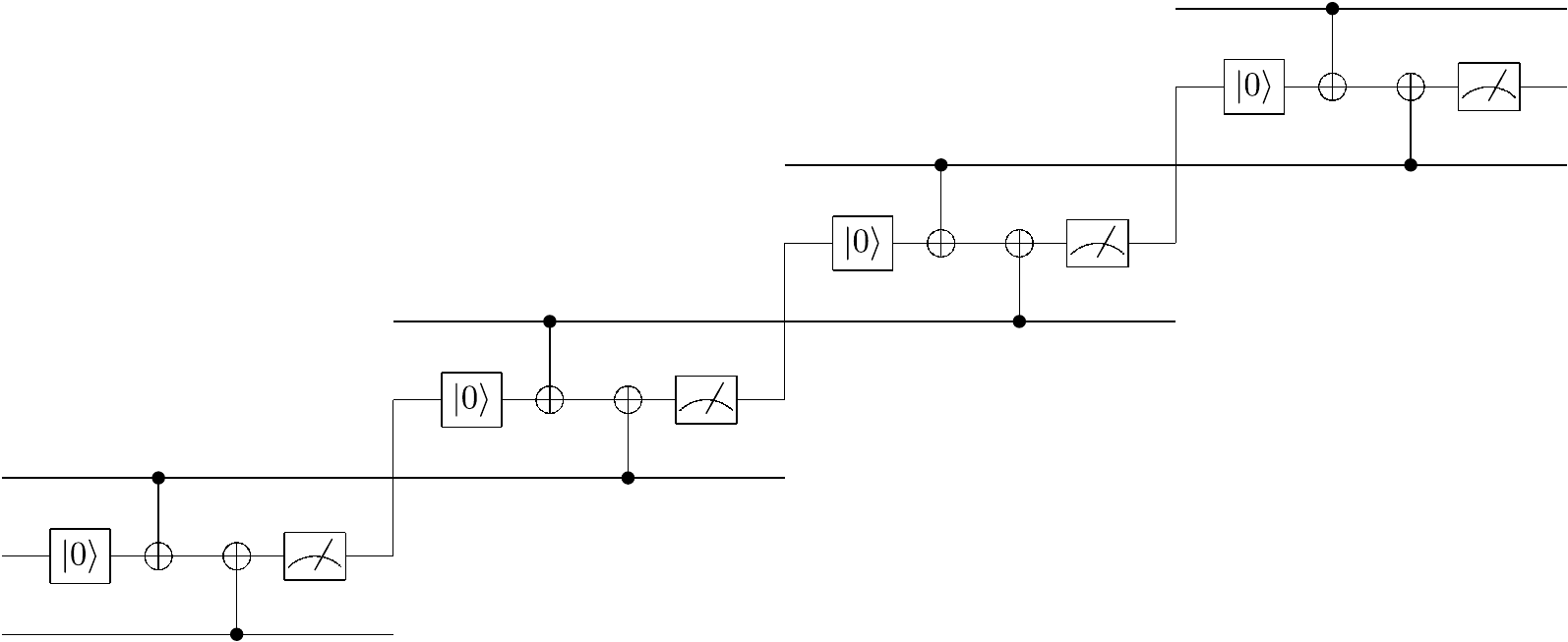}
            }
            \caption{A method to reduce the number of qubits in simulation in the case of the one-dimensional repetition code. (a) The original syndrome measurement circuit. (b) A time-shifted version of the original one.}
            \label{fig:repetition}
        \end{figure}
        If we straightforwardly simulate 49 qubits, we need to reserve a $2^{49}$-dimensional complex vector on a classical memory.
        It prohibits us to simulate the error correction procedure with a practical computational resource.
        To overcome this obstacle, we reduce the number of qubits that have to be simulated by reusing the measured qubits.
        As an illustrative example of this, 
        we show the case for the one-dimensional repetition code in FIG.\ref{fig:repetition}.
        While in the actual experiment, 
        the syndrome measurements are done in parallel for each measurement qubit,
        we delay them so that one syndrome measurement runs at a time.
        This strategy allows us to reuse one measurement qubit for multiple stabiliser measurements without changing the system to be simulated.
        Note that the same effect can be obtained by analytically calculating a set of POVM operators corresponding to each syndrome measurement and applying them with appropriate probabilities. However, the circuit-based noise model with coherent error considered here is somewhat complicated and hence we avoid this approach.

        \begin{figure}[hb]
            \subfloat[\label{subfig:block}]{
                \includegraphics[width=0.6\linewidth]{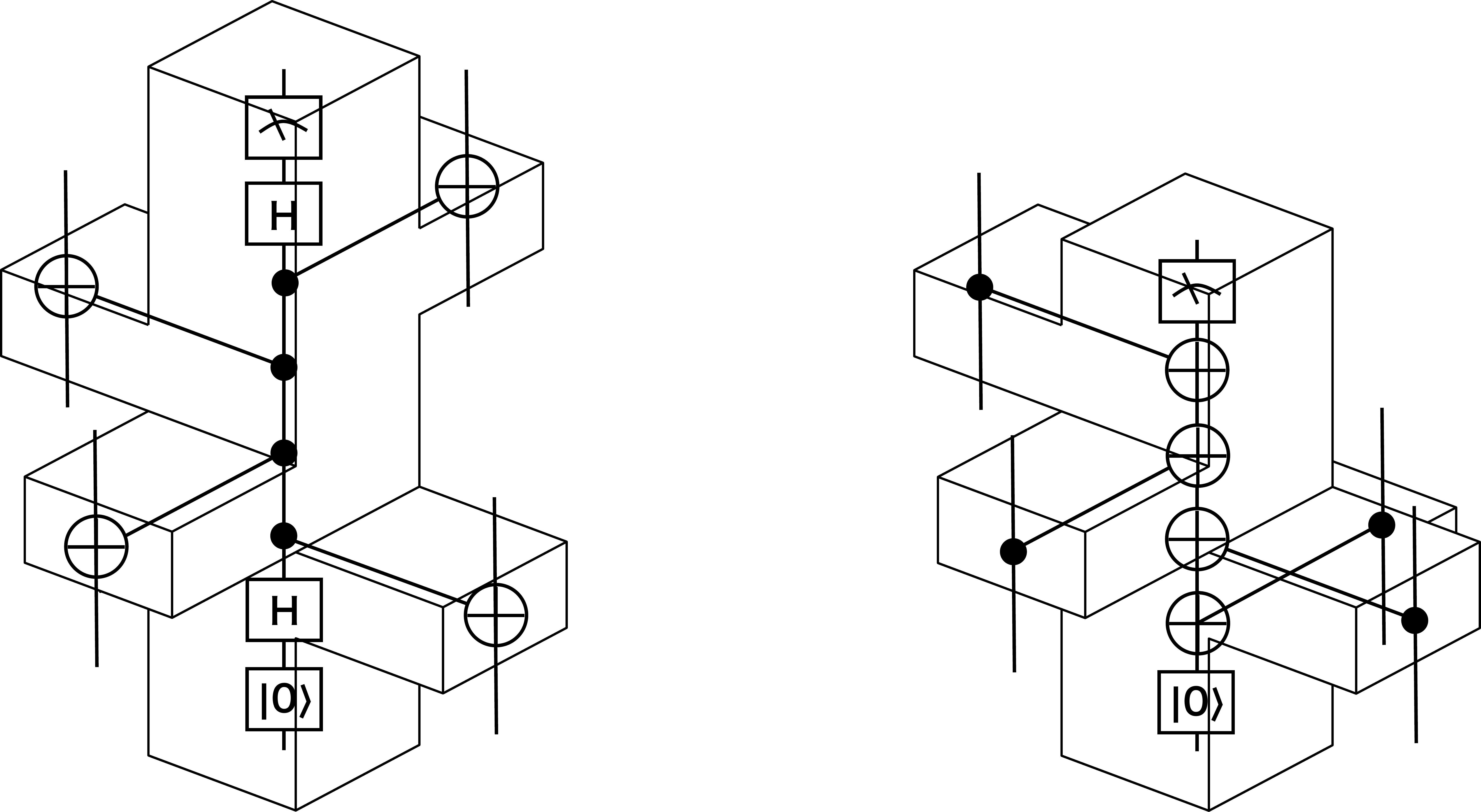}
            }\\
            \subfloat[\label{subfig:blocks}]{
                \includegraphics[width=0.58\linewidth]{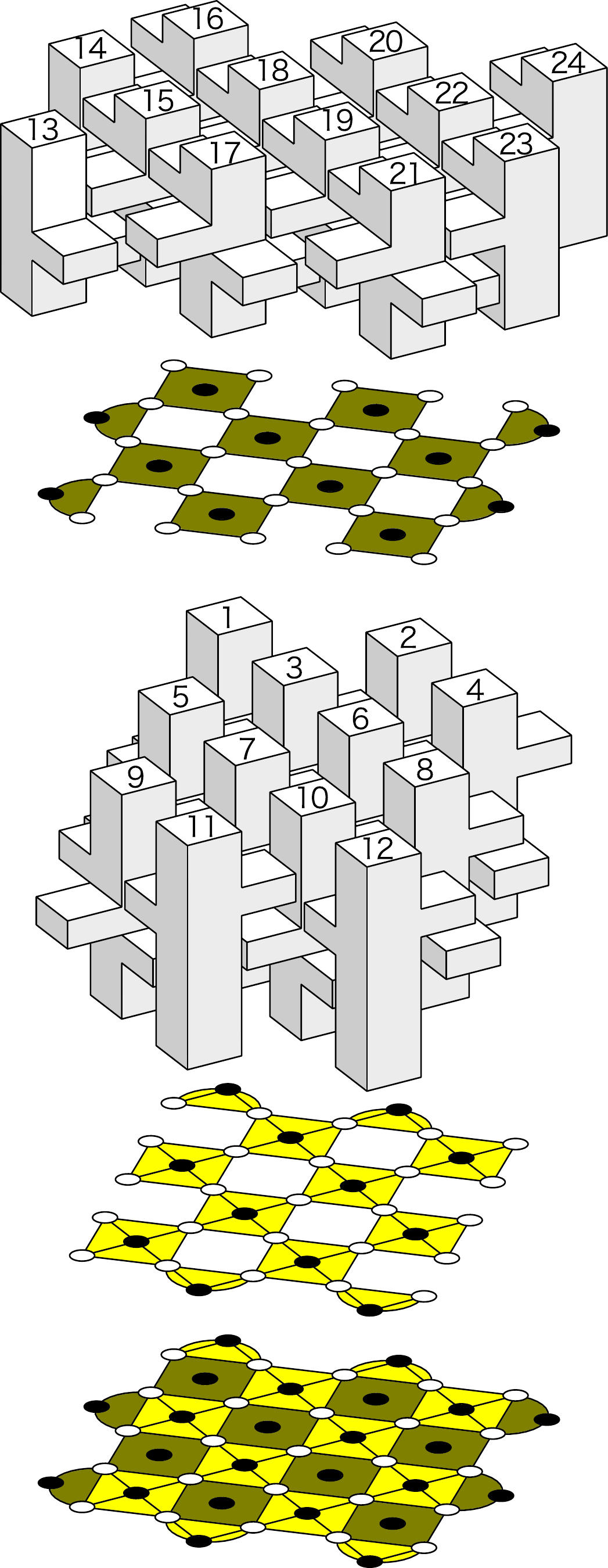}
            }
            \caption{\protect\subref{subfig:block} Left and right blocks represent the syndrome measurement circuits for $X$- and $Z$-stabilizers, respectively. \protect\subref{subfig:blocks} Each syndrome measurement circuit is complied for parallel $X$- and $Z$-type syndrome measurements on the surface codes. Note that each syndrome measurement circuit can be delayed without any collision following the order shown in each block so that the syndrome measurements are done sequentially to reduce the number of qubits in simulation.}
        \end{figure}
        When applying the proposed method to surface codes, we have to take care of the order of the two-qubit gates  in the three-dimensional arrangement
        so that each syndrome measurement circuit can be delayed without any collision.
        A three-dimensional unit block corresponding to the syndrome measurements for $X$- and $Z$-type stabilizers is shown in FIG. \ref{subfig:block}.
        The unit block consists of two types of rectangular blocks stacked on a time axis. 
        The rectangle block of $1\times1\times1$ represents a one-qubit gate applied to a measurement qubit and the rectangle block of $2\times1\times1$ represents a two-qubit gate between a measurement qubits and an adjacent data qubit.
        Using these 3D blocks, we construct a syndrome measurement circuit for the whole surface code as in FIG. \ref{subfig:blocks}.
        As you can see, the blocks are assembled in such a way that they can be delayed without any collision. This indicates that the measurement qubits can be reused as same as the one-dimensional case by delaying the syndrome measurements appropriately.
        Specifically, the numbers written in black circles in FIG. \ref{subfig:rotated-surface-code} correspond to the order in which the measurements are to be executed.
        
\section{Numerical experiment}
    \subsection{Simulated system}
        The simulation method in Sec.~\ref{sec:simulation-method} allows us to effectively perform the full-vector simulation of $d=5$ rotated surface code, and therefore to evaluate its performance under any local noise model.
        We consider a situation where two types of noise, incoherent and coherent, act on physical qubits in a syndrome measurement circuit.
        Here, we describe the concrete gateset and noise model that are used in the simulation.
        
\begin{figure*}
            \centering
            \includegraphics[width=0.9\linewidth]{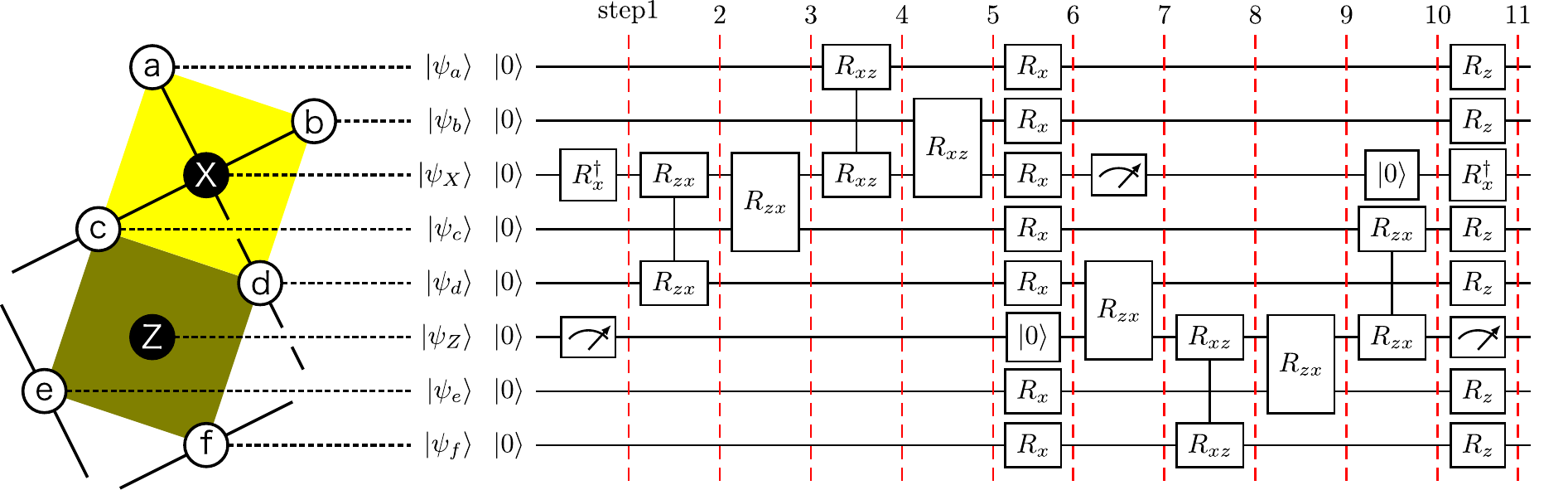}
            \caption{A syndrome measurement circuit for the rotated surface codes consisting of cross resonance gates and Virtual-Z gates. 
            The rotation angle of the $R_x,\,R_z,\,R_{zx}$ gates is $\pi/2$ and that of the $R_x^\dagger$ gate is $-\pi/2$.}
            \label{fig:circuit}
        \end{figure*}

        To conduct a realistic simulation, we consider a hardware-native gateset common to the superconducting devices and compile the syndrome measurement circuit with those gates.
        More concretely, we use a gateset $\{R_x(\pi/2), R_x^\dagger(\pi/2), R_{zx}(\pi/2), R_{zx}^\dagger(\pi/2), R_z(\pi/2)\}$, where
        \begin{align}
            R_x(\pi/2) &= e^{-i\pi X/4}, \\
            R_z(\pi/2) &= e^{-i\pi Z/4}, \\
            R_{zx}(\pi/2) &= e^{-i\pi X\otimes Z/4}, 
        \end{align}
        to perform the syndrome measurements.
        This gateset is commonly used in the superconducting qubits with cross resonance gate \cite{Chow2011, Krantz2019}. 
        With these gate sets, the syndrome measurement circuit can be rewritten as in FIG. \ref{fig:circuit}, where
        three changes are made from the circuit shown in FIG.\ref{subfig:X-stabilizer} and FIG.\ref{subfig:Z-stabilizer}.
        Firstly, we replace CNOT gates in FIG.\ref{subfig:Z-stabilizer} with $R_{zx}(\pi/2)$ gates.
        Second, for the two Hadamard gates in the circuit of FIG. \ref{subfig:X-stabilizer}, we replaced the former one by the $R_x^\dagger(\pi/2)$ gate and the latter one by the $R_x(\pi/2)$ gate.
        Finally, we apply $R_x(\pi/2)$ gates to the data qubits after the circuit shown in FIG.\ref{subfig:X-stabilizer} and $R_z(\pi/2)$ gates to the data qubits after the circuit in FIG.\ref{subfig:Z-stabilizer}.
        These one-qubit rotation gates restore the errors transformed to $Y$ error by $R_{zx}(\pi/2)$ gates.
        
        Next, let us define the noise model used in the simulation and see how incoherent noise is incorporated.
        We assume that the single-qubit depolarizing noise,
        \begin{align}
            \mathcal{E}_1(\rho)&=\qty(1-p)\rho+\frac{p}{3}\sum_{A\in\qty{X,\,Y,\,Z}}A\rho A,
            \label{eq:one-depolarizing-noise}
        \end{align}
        acts on every qubit at each step of the measurement circuit in FIG. \ref{fig:circuit}.
        Same noise occurs in the ``waiting'' qubit where the gate does not act.
        For the two qubits after the cross-resonance gates, we apply the two-qubit depolarizing noise:
        \begin{align}
            \mathcal{E}_2(\rho)
            &=\qty(1-p)\rho+\frac{p}{15}\sum_{A,B\in\qty{I,X,Y,Z}}(A\otimes B)\rho(A\otimes B).
            \label{eq:two-depolarizing-noise}
        \end{align}
        We call the parameter $p$ the physical error probability.
        In this paper, the probability of each Pauli error is set to be equal, but it is not difficult to reflect the actual distribution of stochastic errors.
        
        Next, we describe coherent noise.
        The coherent noise was modelled as an (unintentional) increase in the rotation angle of each rotation gate.
        This increase is referred to as an over-rotation error.
        Since this is associated with the rotation gate, it does not act on the ``waiting'' data qubit.
        In this study, we want to model the error probability with the single parameter $p$.
        To this end, we introduce a coherent error ratio $c$ and set the over-rotation angle to be $\theta=2c\sqrt{p}$.
        In other words, for a rotation gate generated by a Pauli operator $A$, we add the coherent noise channel in the form of:
        \begin{align}
            \mathcal{E}_c(\rho)=e^{-ic\sqrt{p}A}\rho e^{ic\sqrt{p}A}.
        \end{align}
        The reason why we set $\theta=2c\sqrt{p}$ is as follows.
        Consider the expectation value of $P_1=\dyad{1}$ when $R_x(\theta)=e^{-i\frac{\theta}{2}X}$ is applied to $\ket{0}$.
        It is calculated as,
        \begin{align}
            \ev{R_x^\dagger(\theta)P_1R_x(\theta)}{0}=\sin^2\frac{\theta}{2}
        \end{align}
        Since $\theta\ll1$, the bit-flip probability $p_\text{flip}$ associated with this over-rotation is, 
        \begin{align}
            p_\text{flip}&=c^2 p.
        \end{align}
        Or equivalently, if the coherence of noise is destroyed at each step, for example, by using the twirling operation, then such a decohered noise map corresponds to a probabilistic Pauli error with probability $c^2 p$. 
        If the coherent errors experience constructive or destructive interferences, then the effect of the coherent error would be increased or decreased against $c^2 p$.
        The parameter $c$ controls the magnitude of the coherent error compared to the incoherent one.
        
\subsection{Numerical simulation}
        We employ one of the fastest classical simulators of quantum circuits, Qulacs~\cite{suzuki2021qulacs}.
        In the simulation, all data qubits are first initialised to $\ket{0}$ and projected to the surface code state by performing noise-free syndrome measurements with zero outcomes.
        Next, we run syndrome measurement circuits with circuit-level noise as explained above for five rounds.
        After completing five rounds of measurements, all data qubits were subjected to projective measurements in the Pauli $Z$ basis.
        The above procedure was repeated 10000 times.
        In decoding, as usual, we took the XOR of the syndrome of the adjacent rounds to determine the position where the syndrome is flipped.
        With this information, we estimated the error positions using the minimum weight perfect matching (MWPM) algorithm implemented in NetworkX \cite{networkx}.
        After applying the recovery operation using the estimated errors,
         we finally calculate the eigenvalue of the logical Pauli $Z$ operator $Z_L=Z_1Z_7Z_{13}Z_{19}Z_{25}$ from the final projective measurement.
        The logical error probability $p_L$ is estimated by dividing the number of $Z_L=1$ occurrences by 10000.
        We varied $p$ from $10^{-3}$ to $7.0\times10^{-2}$ and $c$ from 0.0 to 1.0 by 0.25.
        For each parameter pair $(p,c)$, the calculation took 18 hours in the absence of coherent errors and up to 80 hours in their presence using an NVIDIA A100 GPU.
        
        \begin{figure}
            \centering
            \includegraphics[width=0.8\linewidth]{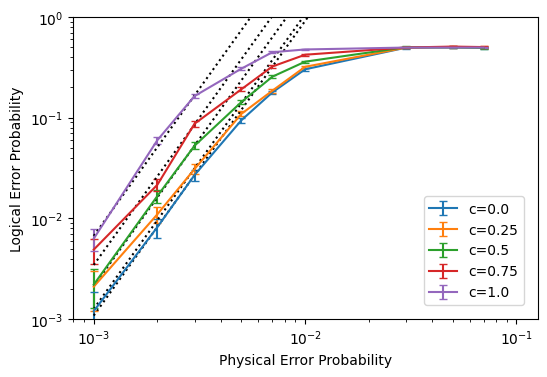}
            \caption{The logical error probability is plotted as a function of the physical error probability $p$ with coherent-noise parameters $c=0,0.25,0.5,0.75,1.0$. The dotted lines correspond to the results of the fitting.}
            \label{fig:pep-lep}
        \end{figure}
        The obtained $p_L$ is shown in FIG. \ref{fig:pep-lep}.
        For $p$ smaller than about $3.0\times10^{-3}$, we can see that $p_L$ decreases if $p$ is decreased. Furthermore, $p_L$ satisfies $p_L=Ap^\xi$ at any value of $c$
        within the statistical error.
        By fitting the numerical values of $p_L$ at $c=0$ for $p=1.0\times 10^{-3}$ to $3.0\times 10^{-3}$ with $p_L=Ap^\xi$ using two parameters $(A,\xi)$, 
        we obtain $A=6.5\times10^5$ and $\xi=2.92$.
        The value of $\xi$ is consistent with the expectation that it should be $\frac{d+1}{2}=3$.
        One might think that the contribution of the coherent error is small because such a small angle rotation can be frozen by repetitive projections for the syndrome measurements.
        However, even for a small ratio $c=0.25$, it has a negligible contribution to the logical error probability.
        This result clearly shows that the effect of the coherence error is important to estimate the experimentally achievable logical error probability accurately.
        
        \subsection{Effective model incorporating the coherent error}
        \begin{figure}
            \centering
            \includegraphics[width=0.8\linewidth]{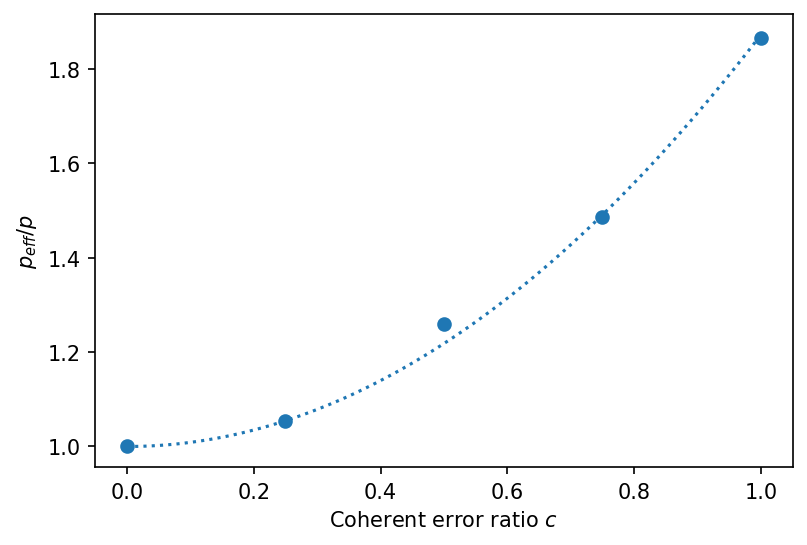}
            \caption{The ratio $p_\mathrm{eff}/p$ is plotted as a function of the coherence parameter $c$. The dotted curve represents $1+\alpha c^2$ with the fitted value $\alpha=0.872$.}
            \label{fig:coefficient-pep}
        \end{figure}
        In order to understand the effect of coherent noise on $p_L$, we consider how to effectively incorporate the effect of coherent error as a leading order correction to the incoherent error $p$.
        Since we expect that the leading-order contribution of the coherent error to the effective error $p_{\mathrm{eff}}$ is proportional to $c^2p$, we model $p_{\mathrm{eff}}$ as,
        \begin{align}
            p_\mathrm{eff}
            =\qty(1+\alpha c^2)p.
            \label{eq:pep-eff}
        \end{align}
        This is because while the probability amplitude for the coherent error is $\sim c\sqrt{p}$, such an error becomes a detectable event if and only if such an error occurs twice on either ket or bra spaces in the density operator picture~\cite{suzuki2017efficient}.
        The coefficient $\alpha$ takes into account the fact that the coherent errors can interfere with each other and therefore their contribution to $p_{\mathrm{eff}}$ can be not exactly $c^2p$. As mentioned before,
        if the coherence is destroyed at each step, then $\alpha$ should be a unit.
        
        Then, we assume that 
        the logical error probability under the coherent error is given by replacing the physical probability $p$ in the case of $c=0$ with $p_{\mathrm{eff}}$.
        More precisely,
        the logical error probability $p_L$ should be obtained by replacing $p$ in Eq. (\ref{eq:pep-eff}) with $p_\mathrm{eff}$:
        \begin{align}
            p_L
            &=A\cdot\qty[\qty(1+\alpha c^2)\cdot p]^{\xi},
            \label{eq:logical-error}
        \end{align}
        where $A$ and $\xi$ are thought to be the same as those with $c=0$.
         The validity of this model is confirmed by detailed numerical calculations carried out on the 1D repetition codes in the previous study \cite{suzuki2017efficient}.
        To test this assumption for the two-dimensional case, we estimate the value of $\alpha$ by the following procedure.
        First, we fit $p_L$ obtained at various $c$'s in the range of $p=1.0\times 10^{-3}$ to $3.0\times 10^{-3}$ with $p_L = A(Bp)^\xi$ using $B$ as the fitting parameter while using the fixed $A$ and $\xi$ obtained at $c=0$.
        Then, we fit $B$ with $1+\alpha c^2$.
        As a result we obtain $\alpha=0.872$ and the ratio $p_{\rm eff}/p$ shown in FIG. \ref{fig:coefficient-pep} implies that this fitting goes well supporting our assumption.
        Another interesting fact is the coefficient $\alpha=0.872$ is smaller than a unit meaning the coherent error interferes in a destructive way. This implies that
        the coherent error modelled in this work is not so damaging for QEC, since the logical error probability is increased when the coherence is destroyed at each step by twirling.
        
        Since this behaviour of the leading order contribution of the coherent error is a local property of the noise, the effective model obtained here is expected to be valid not only when $p$ is reduced for $d=5$, but also when the code distance $d$ is further increased.
        If this is true, we can estimate the logical error probability under the coherent error by combining a limited size of full simulation and a large size of simulation with stochastic Pauli noise.

    \subsection{Experimental consideration}
\begin{figure}
            \begin{tabular}{c}
                \begin{minipage}{1.0\hsize}
                    \centering
                    \includegraphics[width=0.8\linewidth]{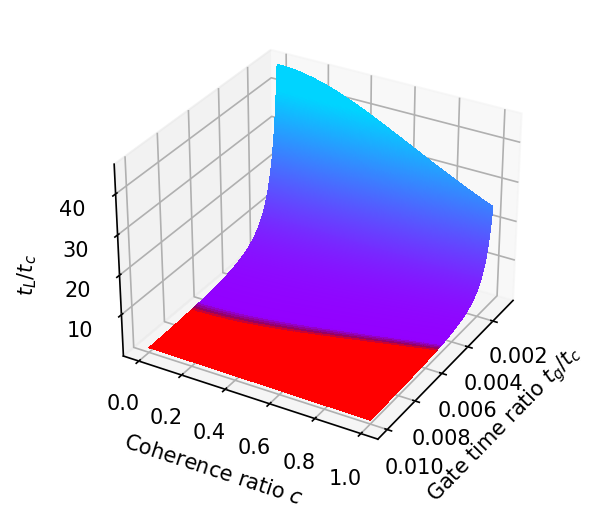}
                \end{minipage}\\
                \begin{minipage}{1.0\hsize}
                    \centering
                    \includegraphics[width=0.8\linewidth]{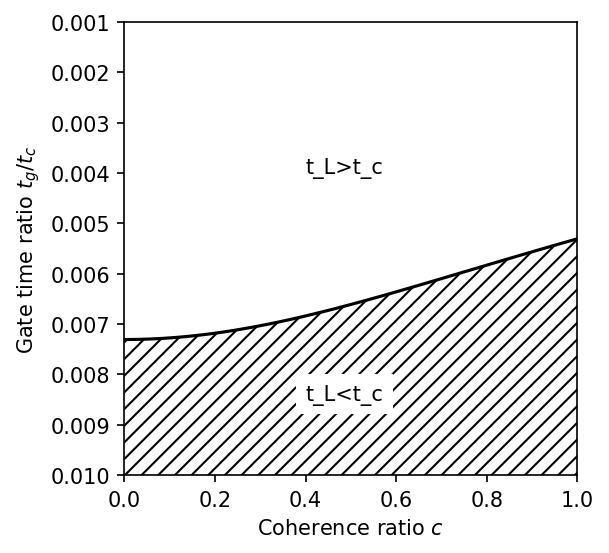}
                \end{minipage}
            \end{tabular}
            \caption{The lifetime $t_L$ of a logical qubit, calculated from the coherent noise parameter $c$, the gate operation time $t_g$ and the coherent time $t_c$ of the physical qubit, divided by $t_c$. The upper graph is a three-dimensional representation of $t_L/t_c$, while the lower graph shows the relationship between $t_g/t_c$ and $c$ in more detail. The red area in the upper graph and the shaded area in the lower graph show the area where $t_L<t_c$.}
            \label{fig:coherent-time}
        \end{figure}

        As a concrete usage of this model,
        here we argue in what situation an experimental QEC can achive a longer lifetime $t_L$ of logical quantum information against a physical coherence time $t_c$ by calculating $t_L/t_c$.
        First, we rewrite the error probability $p$ using 
        the coherence time $t_c$ and time $t_g$ required for each gate operation.
        For clarity, we assume that the error probability is well approximated by:
        \begin{align}
            p \simeq 1-e^{-\frac{t_g}{t_c}}.
            \label{eq:pep}
        \end{align}
        This is the situation where a quantum gate is well-calibrated and the coherence-time-limited fidelity is achieved.
        Second, we rewrite the logical lifetime $t_L$ using the logical error probability $p_L$.
        Letting $N_\text{steps}=11$ be the number of steps in one cycle of the syndrome measurement (see Fig. \ref{fig:circuit}), it takes physical time $d N_\text{steps} t_g$ to conduct $d$ cycles of syndrome measurements.
        Since the logical error occurs with probability $p_L$ in this time, the logical lifetime is roughly given by
        \begin{align}
            t_L
            =\frac{d N_\text{steps} t_g}{p_L}.
            \label{eq:logical-t}
        \end{align}
        Finally, combining Eqs. (\ref{eq:logical-error})-(\ref{eq:logical-t}), we obtain the following relation:
        \begin{align}
            \frac{t_L}{t_c}&=\frac{d\cdot N_\mathrm{gates}}{A\cdot\qty[\beta(1+\alpha c^2)\cdot\qty(1-e^{-\frac{t_g}{t_c}})]^\xi}\frac{t_g}{t_c}.
        \end{align}
        A graph plotting $t_L/t_c$ as a function of $c$ and $t_g/t_c$ is shown in the FIG. \ref{fig:coherent-time}.
        The part of the graph marked in red is the parameter region where $t_L/t_c$ is below 1.
        In this region, the error correction procedure itself damages the lifetime and therefore is meaningless for protecting the quantum information.
        To achieve $t_L/t_c>1$, the gate and coherence time ratio $t_g/t_c$
        depends on the amount of coherent error and ranges from 0.005 to 0.007.
        If the ratio is reduced to $t_g/t_c=0.001$, the lifetime of the logical qubit is improved by a factor of tens against the coherence time even in the presence of the coherence error with $c=1$. 
        In this case, the over-rotation angle is $\theta \sim 0.06$,
        which would be experimentally detectable and hopefully can be calibrated~\cite{sheldon2016characterizing}. 
        
        Finally, let us discuss the experiments on the rotated surface code with code distance 3 carried out by different research groups in 2021~\cite{krinner2022realizing, zhao2022realization}. The main parameters of each experimental system are given in TABLE \ref{tab:parameters}. 
        \begin{table}[b]
            \caption{Parameters of experiments which are implemented by two different groups.}
            \begin{ruledtabular}
                \begin{tabular}{ccc}
                    Parameter & Krinner~\cite{krinner2022realizing} & Zhao~\cite{zhao2022realization}\\
                    \colrule
                    1Q gate duration $T_{g,1}$($\mathrm{ns}$) & 40 & 25\\
                    2Q gate duration $T_{g,2}$($\mathrm{ns}$) & 98 & 32\\
                    Lifetime $T_1$($\mathrm{\mu s}$) & 32.5 & 26.1\\
                    Coherence Time $T_2^\ast$($\mathrm{\mu s}$) & 37.5 & 3.6\\
                    1Q gate error $p_{1Q}$(\%) & 0.09 & 0.098\\
                    2Q gate error $p_{2Q}$(\%) & 1.5 & 1.035\\
                    Measurement duration $T_M$($\mathrm{ns}$) & 300 & 1500\\
                    Measurement error $p_M$(\%) & 0.9 & 4.752\\
                    Logical Lifetime $T_{1,L}$($\mathrm{\mu s}$) & 16.4 & 64.4\\
                    Logical Coherence Time $T_{2,L}^\ast$($\mathrm{\mu s}$) & 18.2 & 69.0
                \end{tabular}
            \end{ruledtabular}
            \label{tab:parameters}
        \end{table}
        For $p_{1Q},\,p_{2Q}$ in the table, Krinner's group evaluated it by Interleaved Randomized Benchmarking (Interleaved RB) and Zhao's group evaluated it by Cross Entropy Benchmarking (XEB).
        Note that while these are great progress toward experimental QEC, neither group has succeeded in QEC in a strict sense.
    	Krinner's group has implemented a QEC protocol, but was unable to make the lifetime of the logical qubit $T_{1, L}$ longer than the lifetime of the physical one $T_{1}$.
        Zhao's group has only implemented the error detection and postselection.
        The main reason for the lack of successful error correction would be the short code distance.
        The distance 3 code cannot correct two-qubit errors that occur during two-qubit gates.

        Let us see whether or not a successful QEC is reachable if the code distance is increased to 5 and consider what elements should be improved if it is not the case.
        Since $p_{2Q}$ is $1.5\,\%,\,1.035\,\%$, which is outside the region where $p_L=Ap^\frac{d+1}{2}$ holds in Figure \ref{fig:pep-lep}, QEC is expected to fail even if the code distance is increased to 5.
    	The ratios of the two-qubit gate and coherence time are
    	$T_{g,2}/T_1=0.003$ and $=0.001$ for Krinners' and Zhaos' groups, respectively. This is sufficiently small from our analysis. However, $p_{2Q}$ are, respectively, given by $1.5\,\%$ and $1.035\,\%$, which is much higher than those limited by the coherence time.
    	This implies that this infidelity caused by a systematic control error or cross-talk resulting in coherent errors. 
        Therefore, by improving their control strategies, there would be a great possibility to achieve a breakeven point by experimental QEC in the near future; at least the coherence time does not become the main obstacle for it.
    
\section{Conclusion}
        In this study, we constructed 
        a framework to fully simulate QEC on the distance 5 rotated surface code under an arbitrary local noise.
        Furthermore, 
        we have constructed an effective model that explains the behaviour of the logical error probability under the coherent errors within the stochastic Pauli noise model with an appropriate modification.
        Therefore, combining our numerical result and the effective model,
        we can analyse the behaviour of the logical error probability with smaller a physical error probability or larger code distance.
        While we only modelled the over-rotation caused by a systematic control error,
        there are plenty of sources of coherent errors 
        such as unintended interactions in Hamiltonian, cross-talk, global fields and so on.
        These sources of noise would be straightforwardly incorporated into our framework.
        The performance analysis under a realistic noise model is becoming increasingly important, and our framework will provide a vital guideline 
        for future improvements on experimental sides.
        
\bibliography{mycite}

\begin{thebibliography}{25}%
\makeatletter
\providecommand \@ifxundefined [1]{%
 \@ifx{#1\undefined}
}%
\providecommand \@ifnum [1]{%
 \ifnum #1\expandafter \@firstoftwo
 \else \expandafter \@secondoftwo
 \fi
}%
\providecommand \@ifx [1]{%
 \ifx #1\expandafter \@firstoftwo
 \else \expandafter \@secondoftwo
 \fi
}%
\providecommand \natexlab [1]{#1}%
\providecommand \enquote  [1]{``#1''}%
\providecommand \bibnamefont  [1]{#1}%
\providecommand \bibfnamefont [1]{#1}%
\providecommand \citenamefont [1]{#1}%
\providecommand \href@noop [0]{\@secondoftwo}%
\providecommand \href [0]{\begingroup \@sanitize@url \@href}%
\providecommand \@href[1]{\@@startlink{#1}\@@href}%
\providecommand \@@href[1]{\endgroup#1\@@endlink}%
\providecommand \@sanitize@url [0]{\catcode `\\12\catcode `\$12\catcode
  `\&12\catcode `\#12\catcode `\^12\catcode `\_12\catcode `\%12\relax}%
\providecommand \@@startlink[1]{}%
\providecommand \@@endlink[0]{}%
\providecommand \url  [0]{\begingroup\@sanitize@url \@url }%
\providecommand \@url [1]{\endgroup\@href {#1}{\urlprefix }}%
\providecommand \urlprefix  [0]{URL }%
\providecommand \Eprint [0]{\href }%
\providecommand \doibase [0]{https://doi.org/}%
\providecommand \selectlanguage [0]{\@gobble}%
\providecommand \bibinfo  [0]{\@secondoftwo}%
\providecommand \bibfield  [0]{\@secondoftwo}%
\providecommand \translation [1]{[#1]}%
\providecommand \BibitemOpen [0]{}%
\providecommand \bibitemStop [0]{}%
\providecommand \bibitemNoStop [0]{.\EOS\space}%
\providecommand \EOS [0]{\spacefactor3000\relax}%
\providecommand \BibitemShut  [1]{\csname bibitem#1\endcsname}%
\let\auto@bib@innerbib\@empty
\bibitem [{\citenamefont {Shor}(1996)}]{shor1996fault}%
  \BibitemOpen
  \bibfield  {author} {\bibinfo {author} {\bibfnamefont {P.}~\bibnamefont
  {Shor}},\ }\bibfield  {title} {\bibinfo {title} {Fault-tolerant quantum
  computation},\ }in\ \href {https://doi.org/10.1109/SFCS.1996.548464} {\emph
  {\bibinfo {booktitle} {Proceedings of 37th Conference on Foundations of
  Computer Science}}}\ (\bibinfo {year} {1996})\ pp.\ \bibinfo {pages}
  {56--65}\BibitemShut {NoStop}%
\bibitem [{\citenamefont {Zhang}\ \emph {et~al.}(2012)\citenamefont {Zhang},
  \citenamefont {Laflamme},\ and\ \citenamefont
  {Suter}}]{zhang2012experimental}%
  \BibitemOpen
  \bibfield  {author} {\bibinfo {author} {\bibfnamefont {J.}~\bibnamefont
  {Zhang}}, \bibinfo {author} {\bibfnamefont {R.}~\bibnamefont {Laflamme}},\
  and\ \bibinfo {author} {\bibfnamefont {D.}~\bibnamefont {Suter}},\ }\bibfield
   {title} {\bibinfo {title} {Experimental implementation of encoded logical
  qubit operations in a perfect quantum error correcting code},\ }\href@noop {}
  {\bibfield  {journal} {\bibinfo  {journal} {Physical review letters}\
  }\textbf {\bibinfo {volume} {109}},\ \bibinfo {pages} {100503} (\bibinfo
  {year} {2012})}\BibitemShut {NoStop}%
\bibitem [{\citenamefont {Abobeih}\ \emph {et~al.}(2022)\citenamefont
  {Abobeih}, \citenamefont {Wang}, \citenamefont {Randall}, \citenamefont
  {Loenen}, \citenamefont {Bradley}, \citenamefont {Markham}, \citenamefont
  {Twitchen}, \citenamefont {Terhal},\ and\ \citenamefont
  {Taminiau}}]{abobeih2021fault}%
  \BibitemOpen
  \bibfield  {author} {\bibinfo {author} {\bibfnamefont {M.}~\bibnamefont
  {Abobeih}}, \bibinfo {author} {\bibfnamefont {Y.}~\bibnamefont {Wang}},
  \bibinfo {author} {\bibfnamefont {J.}~\bibnamefont {Randall}}, \bibinfo
  {author} {\bibfnamefont {S.}~\bibnamefont {Loenen}}, \bibinfo {author}
  {\bibfnamefont {C.}~\bibnamefont {Bradley}}, \bibinfo {author} {\bibfnamefont
  {M.}~\bibnamefont {Markham}}, \bibinfo {author} {\bibfnamefont
  {D.}~\bibnamefont {Twitchen}}, \bibinfo {author} {\bibfnamefont
  {B.}~\bibnamefont {Terhal}},\ and\ \bibinfo {author} {\bibfnamefont
  {T.}~\bibnamefont {Taminiau}},\ }\bibfield  {title} {\bibinfo {title}
  {Fault-tolerant operation of a logical qubit in a diamond quantum
  processor},\ }\href@noop {} {\bibfield  {journal} {\bibinfo  {journal}
  {Nature}\ ,\ \bibinfo {pages} {1}} (\bibinfo {year} {2022})}\BibitemShut
  {NoStop}%
\bibitem [{\citenamefont {Egan}\ \emph {et~al.}(2021)\citenamefont {Egan},
  \citenamefont {Debroy}, \citenamefont {Noel}, \citenamefont {Risinger},
  \citenamefont {Zhu}, \citenamefont {Biswas}, \citenamefont {Newman},
  \citenamefont {Li}, \citenamefont {Brown}, \citenamefont {Cetina} \emph
  {et~al.}}]{egan2021fault}%
  \BibitemOpen
  \bibfield  {author} {\bibinfo {author} {\bibfnamefont {L.}~\bibnamefont
  {Egan}}, \bibinfo {author} {\bibfnamefont {D.~M.}\ \bibnamefont {Debroy}},
  \bibinfo {author} {\bibfnamefont {C.}~\bibnamefont {Noel}}, \bibinfo {author}
  {\bibfnamefont {A.}~\bibnamefont {Risinger}}, \bibinfo {author}
  {\bibfnamefont {D.}~\bibnamefont {Zhu}}, \bibinfo {author} {\bibfnamefont
  {D.}~\bibnamefont {Biswas}}, \bibinfo {author} {\bibfnamefont
  {M.}~\bibnamefont {Newman}}, \bibinfo {author} {\bibfnamefont
  {M.}~\bibnamefont {Li}}, \bibinfo {author} {\bibfnamefont {K.~R.}\
  \bibnamefont {Brown}}, \bibinfo {author} {\bibfnamefont {M.}~\bibnamefont
  {Cetina}}, \emph {et~al.},\ }\bibfield  {title} {\bibinfo {title}
  {Fault-tolerant control of an error-corrected qubit},\ }\href@noop {}
  {\bibfield  {journal} {\bibinfo  {journal} {Nature}\ }\textbf {\bibinfo
  {volume} {598}},\ \bibinfo {pages} {281} (\bibinfo {year}
  {2021})}\BibitemShut {NoStop}%
\bibitem [{\citenamefont {Luo}\ \emph {et~al.}(2021)\citenamefont {Luo},
  \citenamefont {Chen}, \citenamefont {Erhard}, \citenamefont {Zhong},
  \citenamefont {Wu}, \citenamefont {Tang}, \citenamefont {Zhao}, \citenamefont
  {Wang}, \citenamefont {Fujii}, \citenamefont {Li} \emph
  {et~al.}}]{luo2021quantum}%
  \BibitemOpen
  \bibfield  {author} {\bibinfo {author} {\bibfnamefont {Y.-H.}\ \bibnamefont
  {Luo}}, \bibinfo {author} {\bibfnamefont {M.-C.}\ \bibnamefont {Chen}},
  \bibinfo {author} {\bibfnamefont {M.}~\bibnamefont {Erhard}}, \bibinfo
  {author} {\bibfnamefont {H.-S.}\ \bibnamefont {Zhong}}, \bibinfo {author}
  {\bibfnamefont {D.}~\bibnamefont {Wu}}, \bibinfo {author} {\bibfnamefont
  {H.-Y.}\ \bibnamefont {Tang}}, \bibinfo {author} {\bibfnamefont
  {Q.}~\bibnamefont {Zhao}}, \bibinfo {author} {\bibfnamefont {X.-L.}\
  \bibnamefont {Wang}}, \bibinfo {author} {\bibfnamefont {K.}~\bibnamefont
  {Fujii}}, \bibinfo {author} {\bibfnamefont {L.}~\bibnamefont {Li}}, \emph
  {et~al.},\ }\bibfield  {title} {\bibinfo {title} {Quantum teleportation of
  physical qubits into logical code spaces},\ }\href@noop {} {\bibfield
  {journal} {\bibinfo  {journal} {Proceedings of the National Academy of
  Sciences}\ }\textbf {\bibinfo {volume} {118}} (\bibinfo {year}
  {2021})}\BibitemShut {NoStop}%
\bibitem [{\citenamefont {Andersen}\ \emph {et~al.}(2020)\citenamefont
  {Andersen}, \citenamefont {Remm}, \citenamefont {Lazar}, \citenamefont
  {Krinner}, \citenamefont {Lacroix}, \citenamefont {Norris}, \citenamefont
  {Gabureac}, \citenamefont {Eichler},\ and\ \citenamefont
  {Wallraff}}]{andersen2020repeated}%
  \BibitemOpen
  \bibfield  {author} {\bibinfo {author} {\bibfnamefont {C.~K.}\ \bibnamefont
  {Andersen}}, \bibinfo {author} {\bibfnamefont {A.}~\bibnamefont {Remm}},
  \bibinfo {author} {\bibfnamefont {S.}~\bibnamefont {Lazar}}, \bibinfo
  {author} {\bibfnamefont {S.}~\bibnamefont {Krinner}}, \bibinfo {author}
  {\bibfnamefont {N.}~\bibnamefont {Lacroix}}, \bibinfo {author} {\bibfnamefont
  {G.~J.}\ \bibnamefont {Norris}}, \bibinfo {author} {\bibfnamefont
  {M.}~\bibnamefont {Gabureac}}, \bibinfo {author} {\bibfnamefont
  {C.}~\bibnamefont {Eichler}},\ and\ \bibinfo {author} {\bibfnamefont
  {A.}~\bibnamefont {Wallraff}},\ }\bibfield  {title} {\bibinfo {title}
  {Repeated quantum error detection in a surface code},\ }\href@noop {}
  {\bibfield  {journal} {\bibinfo  {journal} {Nature Physics}\ }\textbf
  {\bibinfo {volume} {16}},\ \bibinfo {pages} {875} (\bibinfo {year}
  {2020})}\BibitemShut {NoStop}%
\bibitem [{\citenamefont {AI}(2021)}]{ai2021exponential}%
  \BibitemOpen
  \bibfield  {author} {\bibinfo {author} {\bibfnamefont {G.~Q.}\ \bibnamefont
  {AI}},\ }\bibfield  {title} {\bibinfo {title} {Exponential suppression of bit
  or phase errors with cyclic error correction},\ }\href@noop {} {\bibfield
  {journal} {\bibinfo  {journal} {Nature}\ }\textbf {\bibinfo {volume} {595}},\
  \bibinfo {pages} {383} (\bibinfo {year} {2021})}\BibitemShut {NoStop}%
\bibitem [{\citenamefont {Dennis}\ \emph {et~al.}(2002)\citenamefont {Dennis},
  \citenamefont {Kitaev}, \citenamefont {Landahl},\ and\ \citenamefont
  {Preskill}}]{dennis2002topological}%
  \BibitemOpen
  \bibfield  {author} {\bibinfo {author} {\bibfnamefont {E.}~\bibnamefont
  {Dennis}}, \bibinfo {author} {\bibfnamefont {A.}~\bibnamefont {Kitaev}},
  \bibinfo {author} {\bibfnamefont {A.}~\bibnamefont {Landahl}},\ and\ \bibinfo
  {author} {\bibfnamefont {J.}~\bibnamefont {Preskill}},\ }\bibfield  {title}
  {\bibinfo {title} {Topological quantum memory},\ }\href@noop {} {\bibfield
  {journal} {\bibinfo  {journal} {Journal of Mathematical Physics}\ }\textbf
  {\bibinfo {volume} {43}},\ \bibinfo {pages} {4452} (\bibinfo {year}
  {2002})}\BibitemShut {NoStop}%
\bibitem [{\citenamefont {Horsman}\ \emph {et~al.}(2012)\citenamefont
  {Horsman}, \citenamefont {Fowler}, \citenamefont {Devitt},\ and\
  \citenamefont {Van~Meter}}]{horsman2012surface}%
  \BibitemOpen
  \bibfield  {author} {\bibinfo {author} {\bibfnamefont {C.}~\bibnamefont
  {Horsman}}, \bibinfo {author} {\bibfnamefont {A.~G.}\ \bibnamefont {Fowler}},
  \bibinfo {author} {\bibfnamefont {S.}~\bibnamefont {Devitt}},\ and\ \bibinfo
  {author} {\bibfnamefont {R.}~\bibnamefont {Van~Meter}},\ }\bibfield  {title}
  {\bibinfo {title} {Surface code quantum computing by lattice surgery},\
  }\href@noop {} {\bibfield  {journal} {\bibinfo  {journal} {New Journal of
  Physics}\ }\textbf {\bibinfo {volume} {14}},\ \bibinfo {pages} {123011}
  (\bibinfo {year} {2012})}\BibitemShut {NoStop}%
\bibitem [{\citenamefont {Arute}\ \emph {et~al.}(2019)\citenamefont {Arute},
  \citenamefont {Arya}, \citenamefont {Babbush}, \citenamefont {Bacon},
  \citenamefont {Bardin}, \citenamefont {Barends}, \citenamefont {Biswas},
  \citenamefont {Boixo}, \citenamefont {Brandao}, \citenamefont {Buell},
  \citenamefont {Burkett}, \citenamefont {Chen}, \citenamefont {Chen},
  \citenamefont {Chiaro}, \citenamefont {Collins}, \citenamefont {Courtney},
  \citenamefont {Dunsworth}, \citenamefont {Farhi}, \citenamefont {Foxen},
  \citenamefont {Fowler}, \citenamefont {Gidney}, \citenamefont {Giustina},
  \citenamefont {Graff}, \citenamefont {Guerin}, \citenamefont {Habegger},
  \citenamefont {Harrigan}, \citenamefont {Hartmann}, \citenamefont {Ho},
  \citenamefont {Hoffmann}, \citenamefont {Huang}, \citenamefont {Humble},
  \citenamefont {Isakov}, \citenamefont {Jeffrey}, \citenamefont {Jiang},
  \citenamefont {Kafri}, \citenamefont {Kechedzhi}, \citenamefont {Kelly},
  \citenamefont {Klimov}, \citenamefont {Knysh}, \citenamefont {Korotkov},
  \citenamefont {Kostritsa}, \citenamefont {Landhuis}, \citenamefont
  {Lindmark}, \citenamefont {Lucero}, \citenamefont {Lyakh}, \citenamefont
  {Mandr{\`a}}, \citenamefont {McClean}, \citenamefont {McEwen}, \citenamefont
  {Megrant}, \citenamefont {Mi}, \citenamefont {Michielsen}, \citenamefont
  {Mohseni}, \citenamefont {Mutus}, \citenamefont {Naaman}, \citenamefont
  {Neeley}, \citenamefont {Neill}, \citenamefont {Niu}, \citenamefont {Ostby},
  \citenamefont {Petukhov}, \citenamefont {Platt}, \citenamefont {Quintana},
  \citenamefont {Rieffel}, \citenamefont {Roushan}, \citenamefont {Rubin},
  \citenamefont {Sank}, \citenamefont {Satzinger}, \citenamefont {Smelyanskiy},
  \citenamefont {Sung}, \citenamefont {Trevithick}, \citenamefont
  {Vainsencher}, \citenamefont {Villalonga}, \citenamefont {White},
  \citenamefont {Yao}, \citenamefont {Yeh}, \citenamefont {Zalcman},
  \citenamefont {Neven},\ and\ \citenamefont
  {Martinis}}]{aruteQuantumSupremacyUsing2019}%
  \BibitemOpen
  \bibfield  {author} {\bibinfo {author} {\bibfnamefont {F.}~\bibnamefont
  {Arute}}, \bibinfo {author} {\bibfnamefont {K.}~\bibnamefont {Arya}},
  \bibinfo {author} {\bibfnamefont {R.}~\bibnamefont {Babbush}}, \bibinfo
  {author} {\bibfnamefont {D.}~\bibnamefont {Bacon}}, \bibinfo {author}
  {\bibfnamefont {J.~C.}\ \bibnamefont {Bardin}}, \bibinfo {author}
  {\bibfnamefont {R.}~\bibnamefont {Barends}}, \bibinfo {author} {\bibfnamefont
  {R.}~\bibnamefont {Biswas}}, \bibinfo {author} {\bibfnamefont
  {S.}~\bibnamefont {Boixo}}, \bibinfo {author} {\bibfnamefont {F.~G. S.~L.}\
  \bibnamefont {Brandao}}, \bibinfo {author} {\bibfnamefont {D.~A.}\
  \bibnamefont {Buell}}, \bibinfo {author} {\bibfnamefont {B.}~\bibnamefont
  {Burkett}}, \bibinfo {author} {\bibfnamefont {Y.}~\bibnamefont {Chen}},
  \bibinfo {author} {\bibfnamefont {Z.}~\bibnamefont {Chen}}, \bibinfo {author}
  {\bibfnamefont {B.}~\bibnamefont {Chiaro}}, \bibinfo {author} {\bibfnamefont
  {R.}~\bibnamefont {Collins}}, \bibinfo {author} {\bibfnamefont
  {W.}~\bibnamefont {Courtney}}, \bibinfo {author} {\bibfnamefont
  {A.}~\bibnamefont {Dunsworth}}, \bibinfo {author} {\bibfnamefont
  {E.}~\bibnamefont {Farhi}}, \bibinfo {author} {\bibfnamefont
  {B.}~\bibnamefont {Foxen}}, \bibinfo {author} {\bibfnamefont
  {A.}~\bibnamefont {Fowler}}, \bibinfo {author} {\bibfnamefont
  {C.}~\bibnamefont {Gidney}}, \bibinfo {author} {\bibfnamefont
  {M.}~\bibnamefont {Giustina}}, \bibinfo {author} {\bibfnamefont
  {R.}~\bibnamefont {Graff}}, \bibinfo {author} {\bibfnamefont
  {K.}~\bibnamefont {Guerin}}, \bibinfo {author} {\bibfnamefont
  {S.}~\bibnamefont {Habegger}}, \bibinfo {author} {\bibfnamefont {M.~P.}\
  \bibnamefont {Harrigan}}, \bibinfo {author} {\bibfnamefont {M.~J.}\
  \bibnamefont {Hartmann}}, \bibinfo {author} {\bibfnamefont {A.}~\bibnamefont
  {Ho}}, \bibinfo {author} {\bibfnamefont {M.}~\bibnamefont {Hoffmann}},
  \bibinfo {author} {\bibfnamefont {T.}~\bibnamefont {Huang}}, \bibinfo
  {author} {\bibfnamefont {T.~S.}\ \bibnamefont {Humble}}, \bibinfo {author}
  {\bibfnamefont {S.~V.}\ \bibnamefont {Isakov}}, \bibinfo {author}
  {\bibfnamefont {E.}~\bibnamefont {Jeffrey}}, \bibinfo {author} {\bibfnamefont
  {Z.}~\bibnamefont {Jiang}}, \bibinfo {author} {\bibfnamefont
  {D.}~\bibnamefont {Kafri}}, \bibinfo {author} {\bibfnamefont
  {K.}~\bibnamefont {Kechedzhi}}, \bibinfo {author} {\bibfnamefont
  {J.}~\bibnamefont {Kelly}}, \bibinfo {author} {\bibfnamefont {P.~V.}\
  \bibnamefont {Klimov}}, \bibinfo {author} {\bibfnamefont {S.}~\bibnamefont
  {Knysh}}, \bibinfo {author} {\bibfnamefont {A.}~\bibnamefont {Korotkov}},
  \bibinfo {author} {\bibfnamefont {F.}~\bibnamefont {Kostritsa}}, \bibinfo
  {author} {\bibfnamefont {D.}~\bibnamefont {Landhuis}}, \bibinfo {author}
  {\bibfnamefont {M.}~\bibnamefont {Lindmark}}, \bibinfo {author}
  {\bibfnamefont {E.}~\bibnamefont {Lucero}}, \bibinfo {author} {\bibfnamefont
  {D.}~\bibnamefont {Lyakh}}, \bibinfo {author} {\bibfnamefont
  {S.}~\bibnamefont {Mandr{\`a}}}, \bibinfo {author} {\bibfnamefont {J.~R.}\
  \bibnamefont {McClean}}, \bibinfo {author} {\bibfnamefont {M.}~\bibnamefont
  {McEwen}}, \bibinfo {author} {\bibfnamefont {A.}~\bibnamefont {Megrant}},
  \bibinfo {author} {\bibfnamefont {X.}~\bibnamefont {Mi}}, \bibinfo {author}
  {\bibfnamefont {K.}~\bibnamefont {Michielsen}}, \bibinfo {author}
  {\bibfnamefont {M.}~\bibnamefont {Mohseni}}, \bibinfo {author} {\bibfnamefont
  {J.}~\bibnamefont {Mutus}}, \bibinfo {author} {\bibfnamefont
  {O.}~\bibnamefont {Naaman}}, \bibinfo {author} {\bibfnamefont
  {M.}~\bibnamefont {Neeley}}, \bibinfo {author} {\bibfnamefont
  {C.}~\bibnamefont {Neill}}, \bibinfo {author} {\bibfnamefont {M.~Y.}\
  \bibnamefont {Niu}}, \bibinfo {author} {\bibfnamefont {E.}~\bibnamefont
  {Ostby}}, \bibinfo {author} {\bibfnamefont {A.}~\bibnamefont {Petukhov}},
  \bibinfo {author} {\bibfnamefont {J.~C.}\ \bibnamefont {Platt}}, \bibinfo
  {author} {\bibfnamefont {C.}~\bibnamefont {Quintana}}, \bibinfo {author}
  {\bibfnamefont {E.~G.}\ \bibnamefont {Rieffel}}, \bibinfo {author}
  {\bibfnamefont {P.}~\bibnamefont {Roushan}}, \bibinfo {author} {\bibfnamefont
  {N.~C.}\ \bibnamefont {Rubin}}, \bibinfo {author} {\bibfnamefont
  {D.}~\bibnamefont {Sank}}, \bibinfo {author} {\bibfnamefont {K.~J.}\
  \bibnamefont {Satzinger}}, \bibinfo {author} {\bibfnamefont {V.}~\bibnamefont
  {Smelyanskiy}}, \bibinfo {author} {\bibfnamefont {K.~J.}\ \bibnamefont
  {Sung}}, \bibinfo {author} {\bibfnamefont {M.~D.}\ \bibnamefont
  {Trevithick}}, \bibinfo {author} {\bibfnamefont {A.}~\bibnamefont
  {Vainsencher}}, \bibinfo {author} {\bibfnamefont {B.}~\bibnamefont
  {Villalonga}}, \bibinfo {author} {\bibfnamefont {T.}~\bibnamefont {White}},
  \bibinfo {author} {\bibfnamefont {Z.~J.}\ \bibnamefont {Yao}}, \bibinfo
  {author} {\bibfnamefont {P.}~\bibnamefont {Yeh}}, \bibinfo {author}
  {\bibfnamefont {A.}~\bibnamefont {Zalcman}}, \bibinfo {author} {\bibfnamefont
  {H.}~\bibnamefont {Neven}},\ and\ \bibinfo {author} {\bibfnamefont {J.~M.}\
  \bibnamefont {Martinis}},\ }\bibfield  {title} {\bibinfo {title} {Quantum
  supremacy using a programmable superconducting processor},\ }\href
  {https://doi.org/10.1038/s41586-019-1666-5} {\bibfield  {journal} {\bibinfo
  {journal} {Nature}\ }\textbf {\bibinfo {volume} {574}},\ \bibinfo {pages}
  {505} (\bibinfo {year} {2019})}\BibitemShut {NoStop}%
\bibitem [{\citenamefont {Wu}\ \emph {et~al.}(2021)\citenamefont {Wu},
  \citenamefont {Bao}, \citenamefont {Cao}, \citenamefont {Chen}, \citenamefont
  {Chen}, \citenamefont {Chen}, \citenamefont {Chung}, \citenamefont {Deng},
  \citenamefont {Du}, \citenamefont {Fan}, \citenamefont {Gong}, \citenamefont
  {Guo}, \citenamefont {Guo}, \citenamefont {Guo}, \citenamefont {Han},
  \citenamefont {Hong}, \citenamefont {Huang}, \citenamefont {Huo},
  \citenamefont {Li}, \citenamefont {Li}, \citenamefont {Li}, \citenamefont
  {Li}, \citenamefont {Liang}, \citenamefont {Lin}, \citenamefont {Lin},
  \citenamefont {Qian}, \citenamefont {Qiao}, \citenamefont {Rong},
  \citenamefont {Su}, \citenamefont {Sun}, \citenamefont {Wang}, \citenamefont
  {Wang}, \citenamefont {Wu}, \citenamefont {Xu}, \citenamefont {Yan},
  \citenamefont {Yang}, \citenamefont {Yang}, \citenamefont {Ye}, \citenamefont
  {Yin}, \citenamefont {Ying}, \citenamefont {Yu}, \citenamefont {Zha},
  \citenamefont {Zhang}, \citenamefont {Zhang}, \citenamefont {Zhang},
  \citenamefont {Zhang}, \citenamefont {Zhao}, \citenamefont {Zhao},
  \citenamefont {Zhou}, \citenamefont {Zhu}, \citenamefont {Lu}, \citenamefont
  {Peng}, \citenamefont {Zhu},\ and\ \citenamefont {Pan}}]{Wu2021}%
  \BibitemOpen
  \bibfield  {author} {\bibinfo {author} {\bibfnamefont {Y.}~\bibnamefont
  {Wu}}, \bibinfo {author} {\bibfnamefont {W.-S.}\ \bibnamefont {Bao}},
  \bibinfo {author} {\bibfnamefont {S.}~\bibnamefont {Cao}}, \bibinfo {author}
  {\bibfnamefont {F.}~\bibnamefont {Chen}}, \bibinfo {author} {\bibfnamefont
  {M.-C.}\ \bibnamefont {Chen}}, \bibinfo {author} {\bibfnamefont
  {X.}~\bibnamefont {Chen}}, \bibinfo {author} {\bibfnamefont {T.-H.}\
  \bibnamefont {Chung}}, \bibinfo {author} {\bibfnamefont {H.}~\bibnamefont
  {Deng}}, \bibinfo {author} {\bibfnamefont {Y.}~\bibnamefont {Du}}, \bibinfo
  {author} {\bibfnamefont {D.}~\bibnamefont {Fan}}, \bibinfo {author}
  {\bibfnamefont {M.}~\bibnamefont {Gong}}, \bibinfo {author} {\bibfnamefont
  {C.}~\bibnamefont {Guo}}, \bibinfo {author} {\bibfnamefont {C.}~\bibnamefont
  {Guo}}, \bibinfo {author} {\bibfnamefont {S.}~\bibnamefont {Guo}}, \bibinfo
  {author} {\bibfnamefont {L.}~\bibnamefont {Han}}, \bibinfo {author}
  {\bibfnamefont {L.}~\bibnamefont {Hong}}, \bibinfo {author} {\bibfnamefont
  {H.-L.}\ \bibnamefont {Huang}}, \bibinfo {author} {\bibfnamefont {Y.-H.}\
  \bibnamefont {Huo}}, \bibinfo {author} {\bibfnamefont {L.}~\bibnamefont
  {Li}}, \bibinfo {author} {\bibfnamefont {N.}~\bibnamefont {Li}}, \bibinfo
  {author} {\bibfnamefont {S.}~\bibnamefont {Li}}, \bibinfo {author}
  {\bibfnamefont {Y.}~\bibnamefont {Li}}, \bibinfo {author} {\bibfnamefont
  {F.}~\bibnamefont {Liang}}, \bibinfo {author} {\bibfnamefont
  {C.}~\bibnamefont {Lin}}, \bibinfo {author} {\bibfnamefont {J.}~\bibnamefont
  {Lin}}, \bibinfo {author} {\bibfnamefont {H.}~\bibnamefont {Qian}}, \bibinfo
  {author} {\bibfnamefont {D.}~\bibnamefont {Qiao}}, \bibinfo {author}
  {\bibfnamefont {H.}~\bibnamefont {Rong}}, \bibinfo {author} {\bibfnamefont
  {H.}~\bibnamefont {Su}}, \bibinfo {author} {\bibfnamefont {L.}~\bibnamefont
  {Sun}}, \bibinfo {author} {\bibfnamefont {L.}~\bibnamefont {Wang}}, \bibinfo
  {author} {\bibfnamefont {S.}~\bibnamefont {Wang}}, \bibinfo {author}
  {\bibfnamefont {D.}~\bibnamefont {Wu}}, \bibinfo {author} {\bibfnamefont
  {Y.}~\bibnamefont {Xu}}, \bibinfo {author} {\bibfnamefont {K.}~\bibnamefont
  {Yan}}, \bibinfo {author} {\bibfnamefont {W.}~\bibnamefont {Yang}}, \bibinfo
  {author} {\bibfnamefont {Y.}~\bibnamefont {Yang}}, \bibinfo {author}
  {\bibfnamefont {Y.}~\bibnamefont {Ye}}, \bibinfo {author} {\bibfnamefont
  {J.}~\bibnamefont {Yin}}, \bibinfo {author} {\bibfnamefont {C.}~\bibnamefont
  {Ying}}, \bibinfo {author} {\bibfnamefont {J.}~\bibnamefont {Yu}}, \bibinfo
  {author} {\bibfnamefont {C.}~\bibnamefont {Zha}}, \bibinfo {author}
  {\bibfnamefont {C.}~\bibnamefont {Zhang}}, \bibinfo {author} {\bibfnamefont
  {H.}~\bibnamefont {Zhang}}, \bibinfo {author} {\bibfnamefont
  {K.}~\bibnamefont {Zhang}}, \bibinfo {author} {\bibfnamefont
  {Y.}~\bibnamefont {Zhang}}, \bibinfo {author} {\bibfnamefont
  {H.}~\bibnamefont {Zhao}}, \bibinfo {author} {\bibfnamefont {Y.}~\bibnamefont
  {Zhao}}, \bibinfo {author} {\bibfnamefont {L.}~\bibnamefont {Zhou}}, \bibinfo
  {author} {\bibfnamefont {Q.}~\bibnamefont {Zhu}}, \bibinfo {author}
  {\bibfnamefont {C.-Y.}\ \bibnamefont {Lu}}, \bibinfo {author} {\bibfnamefont
  {C.-Z.}\ \bibnamefont {Peng}}, \bibinfo {author} {\bibfnamefont
  {X.}~\bibnamefont {Zhu}},\ and\ \bibinfo {author} {\bibfnamefont {J.-W.}\
  \bibnamefont {Pan}},\ }\bibfield  {title} {\bibinfo {title} {Strong quantum
  computational advantage using a superconducting quantum processor},\ }\href
  {https://doi.org/10.1103/PhysRevLett.127.180501} {\bibfield  {journal}
  {\bibinfo  {journal} {Phys. Rev. Lett.}\ }\textbf {\bibinfo {volume} {127}},\
  \bibinfo {pages} {180501} (\bibinfo {year} {2021})}\BibitemShut {NoStop}%
\bibitem [{\citenamefont {Krinner}\ \emph {et~al.}(2022)\citenamefont
  {Krinner}, \citenamefont {Lacroix}, \citenamefont {Remm}, \citenamefont
  {Di~Paolo}, \citenamefont {Genois}, \citenamefont {Leroux}, \citenamefont
  {Hellings}, \citenamefont {Lazar}, \citenamefont {Swiadek}, \citenamefont
  {Herrmann} \emph {et~al.}}]{krinner2022realizing}%
  \BibitemOpen
  \bibfield  {author} {\bibinfo {author} {\bibfnamefont {S.}~\bibnamefont
  {Krinner}}, \bibinfo {author} {\bibfnamefont {N.}~\bibnamefont {Lacroix}},
  \bibinfo {author} {\bibfnamefont {A.}~\bibnamefont {Remm}}, \bibinfo {author}
  {\bibfnamefont {A.}~\bibnamefont {Di~Paolo}}, \bibinfo {author}
  {\bibfnamefont {E.}~\bibnamefont {Genois}}, \bibinfo {author} {\bibfnamefont
  {C.}~\bibnamefont {Leroux}}, \bibinfo {author} {\bibfnamefont
  {C.}~\bibnamefont {Hellings}}, \bibinfo {author} {\bibfnamefont
  {S.}~\bibnamefont {Lazar}}, \bibinfo {author} {\bibfnamefont
  {F.}~\bibnamefont {Swiadek}}, \bibinfo {author} {\bibfnamefont
  {J.}~\bibnamefont {Herrmann}}, \emph {et~al.},\ }\bibfield  {title} {\bibinfo
  {title} {Realizing repeated quantum error correction in a distance-three
  surface code},\ }\href@noop {} {\bibfield  {journal} {\bibinfo  {journal}
  {Nature}\ }\textbf {\bibinfo {volume} {605}},\ \bibinfo {pages} {669}
  (\bibinfo {year} {2022})}\BibitemShut {NoStop}%
\bibitem [{\citenamefont {Zhao}\ \emph {et~al.}(2022)\citenamefont {Zhao},
  \citenamefont {Ye}, \citenamefont {Huang}, \citenamefont {Zhang},
  \citenamefont {Wu}, \citenamefont {Guan}, \citenamefont {Zhu}, \citenamefont
  {Wei}, \citenamefont {He}, \citenamefont {Cao}, \citenamefont {Chen},
  \citenamefont {Chung}, \citenamefont {Deng}, \citenamefont {Fan},
  \citenamefont {Gong}, \citenamefont {Guo}, \citenamefont {Guo}, \citenamefont
  {Han}, \citenamefont {Li}, \citenamefont {Li}, \citenamefont {Li},
  \citenamefont {Liang}, \citenamefont {Lin}, \citenamefont {Qian},
  \citenamefont {Rong}, \citenamefont {Su}, \citenamefont {Sun}, \citenamefont
  {Wang}, \citenamefont {Wu}, \citenamefont {Xu}, \citenamefont {Ying},
  \citenamefont {Yu}, \citenamefont {Zha}, \citenamefont {Zhang}, \citenamefont
  {Huo}, \citenamefont {Lu}, \citenamefont {Peng}, \citenamefont {Zhu},\ and\
  \citenamefont {Pan}}]{zhao2022realization}%
  \BibitemOpen
  \bibfield  {author} {\bibinfo {author} {\bibfnamefont {Y.}~\bibnamefont
  {Zhao}}, \bibinfo {author} {\bibfnamefont {Y.}~\bibnamefont {Ye}}, \bibinfo
  {author} {\bibfnamefont {H.-L.}\ \bibnamefont {Huang}}, \bibinfo {author}
  {\bibfnamefont {Y.}~\bibnamefont {Zhang}}, \bibinfo {author} {\bibfnamefont
  {D.}~\bibnamefont {Wu}}, \bibinfo {author} {\bibfnamefont {H.}~\bibnamefont
  {Guan}}, \bibinfo {author} {\bibfnamefont {Q.}~\bibnamefont {Zhu}}, \bibinfo
  {author} {\bibfnamefont {Z.}~\bibnamefont {Wei}}, \bibinfo {author}
  {\bibfnamefont {T.}~\bibnamefont {He}}, \bibinfo {author} {\bibfnamefont
  {S.}~\bibnamefont {Cao}}, \bibinfo {author} {\bibfnamefont {F.}~\bibnamefont
  {Chen}}, \bibinfo {author} {\bibfnamefont {T.-H.}\ \bibnamefont {Chung}},
  \bibinfo {author} {\bibfnamefont {H.}~\bibnamefont {Deng}}, \bibinfo {author}
  {\bibfnamefont {D.}~\bibnamefont {Fan}}, \bibinfo {author} {\bibfnamefont
  {M.}~\bibnamefont {Gong}}, \bibinfo {author} {\bibfnamefont {C.}~\bibnamefont
  {Guo}}, \bibinfo {author} {\bibfnamefont {S.}~\bibnamefont {Guo}}, \bibinfo
  {author} {\bibfnamefont {L.}~\bibnamefont {Han}}, \bibinfo {author}
  {\bibfnamefont {N.}~\bibnamefont {Li}}, \bibinfo {author} {\bibfnamefont
  {S.}~\bibnamefont {Li}}, \bibinfo {author} {\bibfnamefont {Y.}~\bibnamefont
  {Li}}, \bibinfo {author} {\bibfnamefont {F.}~\bibnamefont {Liang}}, \bibinfo
  {author} {\bibfnamefont {J.}~\bibnamefont {Lin}}, \bibinfo {author}
  {\bibfnamefont {H.}~\bibnamefont {Qian}}, \bibinfo {author} {\bibfnamefont
  {H.}~\bibnamefont {Rong}}, \bibinfo {author} {\bibfnamefont {H.}~\bibnamefont
  {Su}}, \bibinfo {author} {\bibfnamefont {L.}~\bibnamefont {Sun}}, \bibinfo
  {author} {\bibfnamefont {S.}~\bibnamefont {Wang}}, \bibinfo {author}
  {\bibfnamefont {Y.}~\bibnamefont {Wu}}, \bibinfo {author} {\bibfnamefont
  {Y.}~\bibnamefont {Xu}}, \bibinfo {author} {\bibfnamefont {C.}~\bibnamefont
  {Ying}}, \bibinfo {author} {\bibfnamefont {J.}~\bibnamefont {Yu}}, \bibinfo
  {author} {\bibfnamefont {C.}~\bibnamefont {Zha}}, \bibinfo {author}
  {\bibfnamefont {K.}~\bibnamefont {Zhang}}, \bibinfo {author} {\bibfnamefont
  {Y.-H.}\ \bibnamefont {Huo}}, \bibinfo {author} {\bibfnamefont {C.-Y.}\
  \bibnamefont {Lu}}, \bibinfo {author} {\bibfnamefont {C.-Z.}\ \bibnamefont
  {Peng}}, \bibinfo {author} {\bibfnamefont {X.}~\bibnamefont {Zhu}},\ and\
  \bibinfo {author} {\bibfnamefont {J.-W.}\ \bibnamefont {Pan}},\ }\bibfield
  {title} {\bibinfo {title} {Realization of an error-correcting surface code
  with superconducting qubits},\ }\href
  {https://doi.org/10.1103/PhysRevLett.129.030501} {\bibfield  {journal}
  {\bibinfo  {journal} {Phys. Rev. Lett.}\ }\textbf {\bibinfo {volume} {129}},\
  \bibinfo {pages} {030501} (\bibinfo {year} {2022})}\BibitemShut {NoStop}%
\bibitem [{\citenamefont {Bluvstein}\ \emph {et~al.}(2022)\citenamefont
  {Bluvstein}, \citenamefont {Levine}, \citenamefont {Semeghini}, \citenamefont
  {Wang}, \citenamefont {Ebadi}, \citenamefont {Kalinowski}, \citenamefont
  {Keesling}, \citenamefont {Maskara}, \citenamefont {Pichler}, \citenamefont
  {Greiner} \emph {et~al.}}]{Bluvstein2021}%
  \BibitemOpen
  \bibfield  {author} {\bibinfo {author} {\bibfnamefont {D.}~\bibnamefont
  {Bluvstein}}, \bibinfo {author} {\bibfnamefont {H.}~\bibnamefont {Levine}},
  \bibinfo {author} {\bibfnamefont {G.}~\bibnamefont {Semeghini}}, \bibinfo
  {author} {\bibfnamefont {T.~T.}\ \bibnamefont {Wang}}, \bibinfo {author}
  {\bibfnamefont {S.}~\bibnamefont {Ebadi}}, \bibinfo {author} {\bibfnamefont
  {M.}~\bibnamefont {Kalinowski}}, \bibinfo {author} {\bibfnamefont
  {A.}~\bibnamefont {Keesling}}, \bibinfo {author} {\bibfnamefont
  {N.}~\bibnamefont {Maskara}}, \bibinfo {author} {\bibfnamefont
  {H.}~\bibnamefont {Pichler}}, \bibinfo {author} {\bibfnamefont
  {M.}~\bibnamefont {Greiner}}, \emph {et~al.},\ }\bibfield  {title} {\bibinfo
  {title} {A quantum processor based on coherent transport of entangled atom
  arrays},\ }\href@noop {} {\bibfield  {journal} {\bibinfo  {journal} {Nature}\
  }\textbf {\bibinfo {volume} {604}},\ \bibinfo {pages} {451} (\bibinfo {year}
  {2022})}\BibitemShut {NoStop}%
\bibitem [{\citenamefont {C{\'o}rcoles}\ \emph {et~al.}(2015)\citenamefont
  {C{\'o}rcoles}, \citenamefont {Magesan}, \citenamefont {Srinivasan},
  \citenamefont {Cross}, \citenamefont {Steffen}, \citenamefont {Gambetta},\
  and\ \citenamefont {Chow}}]{corcoles2015demonstration}%
  \BibitemOpen
  \bibfield  {author} {\bibinfo {author} {\bibfnamefont {A.~D.}\ \bibnamefont
  {C{\'o}rcoles}}, \bibinfo {author} {\bibfnamefont {E.}~\bibnamefont
  {Magesan}}, \bibinfo {author} {\bibfnamefont {S.~J.}\ \bibnamefont
  {Srinivasan}}, \bibinfo {author} {\bibfnamefont {A.~W.}\ \bibnamefont
  {Cross}}, \bibinfo {author} {\bibfnamefont {M.}~\bibnamefont {Steffen}},
  \bibinfo {author} {\bibfnamefont {J.~M.}\ \bibnamefont {Gambetta}},\ and\
  \bibinfo {author} {\bibfnamefont {J.~M.}\ \bibnamefont {Chow}},\ }\bibfield
  {title} {\bibinfo {title} {Demonstration of a quantum error detection code
  using a square lattice of four superconducting qubits},\ }\href@noop {}
  {\bibfield  {journal} {\bibinfo  {journal} {Nature communications}\ }\textbf
  {\bibinfo {volume} {6}},\ \bibinfo {pages} {1} (\bibinfo {year}
  {2015})}\BibitemShut {NoStop}%
\bibitem [{\citenamefont {Nielsen}\ and\ \citenamefont
  {Chuang}(2010)}]{nielsen2010quantum}%
  \BibitemOpen
  \bibfield  {author} {\bibinfo {author} {\bibfnamefont {M.~A.}\ \bibnamefont
  {Nielsen}}\ and\ \bibinfo {author} {\bibfnamefont {I.~L.}\ \bibnamefont
  {Chuang}},\ }\href {https://doi.org/10.1017/CBO9780511976667} {\emph
  {\bibinfo {title} {Quantum Computation and Quantum Information: 10th
  Anniversary Edition}}}\ (\bibinfo  {publisher} {Cambridge University Press},\
  \bibinfo {year} {2010})\BibitemShut {NoStop}%
\bibitem [{\citenamefont {Suzuki}\ \emph {et~al.}(2017)\citenamefont {Suzuki},
  \citenamefont {Fujii},\ and\ \citenamefont {Koashi}}]{suzuki2017efficient}%
  \BibitemOpen
  \bibfield  {author} {\bibinfo {author} {\bibfnamefont {Y.}~\bibnamefont
  {Suzuki}}, \bibinfo {author} {\bibfnamefont {K.}~\bibnamefont {Fujii}},\ and\
  \bibinfo {author} {\bibfnamefont {M.}~\bibnamefont {Koashi}},\ }\bibfield
  {title} {\bibinfo {title} {Efficient simulation of quantum error correction
  under coherent error based on the nonunitary free-fermionic formalism},\
  }\href@noop {} {\bibfield  {journal} {\bibinfo  {journal} {Physical review
  letters}\ }\textbf {\bibinfo {volume} {119}},\ \bibinfo {pages} {190503}
  (\bibinfo {year} {2017})}\BibitemShut {NoStop}%
\bibitem [{\citenamefont {Tomita}\ and\ \citenamefont
  {Svore}(2014)}]{tomita2014low}%
  \BibitemOpen
  \bibfield  {author} {\bibinfo {author} {\bibfnamefont {Y.}~\bibnamefont
  {Tomita}}\ and\ \bibinfo {author} {\bibfnamefont {K.~M.}\ \bibnamefont
  {Svore}},\ }\bibfield  {title} {\bibinfo {title} {Low-distance surface codes
  under realistic quantum noise},\ }\href@noop {} {\bibfield  {journal}
  {\bibinfo  {journal} {Physical Review A}\ }\textbf {\bibinfo {volume} {90}},\
  \bibinfo {pages} {062320} (\bibinfo {year} {2014})}\BibitemShut {NoStop}%
\bibitem [{\citenamefont {Darmawan}\ and\ \citenamefont
  {Poulin}(2017)}]{darmawan2017tensor}%
  \BibitemOpen
  \bibfield  {author} {\bibinfo {author} {\bibfnamefont {A.~S.}\ \bibnamefont
  {Darmawan}}\ and\ \bibinfo {author} {\bibfnamefont {D.}~\bibnamefont
  {Poulin}},\ }\bibfield  {title} {\bibinfo {title} {Tensor-network simulations
  of the surface code under realistic noise},\ }\href@noop {} {\bibfield
  {journal} {\bibinfo  {journal} {Physical review letters}\ }\textbf {\bibinfo
  {volume} {119}},\ \bibinfo {pages} {040502} (\bibinfo {year}
  {2017})}\BibitemShut {NoStop}%
\bibitem [{\citenamefont {Hakkaku}\ \emph {et~al.}(2021)\citenamefont
  {Hakkaku}, \citenamefont {Mitarai},\ and\ \citenamefont
  {Fujii}}]{hakkaku2021sampling}%
  \BibitemOpen
  \bibfield  {author} {\bibinfo {author} {\bibfnamefont {S.}~\bibnamefont
  {Hakkaku}}, \bibinfo {author} {\bibfnamefont {K.}~\bibnamefont {Mitarai}},\
  and\ \bibinfo {author} {\bibfnamefont {K.}~\bibnamefont {Fujii}},\ }\bibfield
   {title} {\bibinfo {title} {Sampling-based quasiprobability simulation for
  fault-tolerant quantum error correction on the surface codes under coherent
  noise},\ }\href {https://doi.org/10.1103/PhysRevResearch.3.043130} {\bibfield
   {journal} {\bibinfo  {journal} {Phys. Rev. Research}\ }\textbf {\bibinfo
  {volume} {3}},\ \bibinfo {pages} {043130} (\bibinfo {year}
  {2021})}\BibitemShut {NoStop}%
\bibitem [{\citenamefont {Chow}\ \emph {et~al.}(2011)\citenamefont {Chow},
  \citenamefont {C\'orcoles}, \citenamefont {Gambetta}, \citenamefont
  {Rigetti}, \citenamefont {Johnson}, \citenamefont {Smolin}, \citenamefont
  {Rozen}, \citenamefont {Keefe}, \citenamefont {Rothwell}, \citenamefont
  {Ketchen},\ and\ \citenamefont {Steffen}}]{Chow2011}%
  \BibitemOpen
  \bibfield  {author} {\bibinfo {author} {\bibfnamefont {J.~M.}\ \bibnamefont
  {Chow}}, \bibinfo {author} {\bibfnamefont {A.~D.}\ \bibnamefont
  {C\'orcoles}}, \bibinfo {author} {\bibfnamefont {J.~M.}\ \bibnamefont
  {Gambetta}}, \bibinfo {author} {\bibfnamefont {C.}~\bibnamefont {Rigetti}},
  \bibinfo {author} {\bibfnamefont {B.~R.}\ \bibnamefont {Johnson}}, \bibinfo
  {author} {\bibfnamefont {J.~A.}\ \bibnamefont {Smolin}}, \bibinfo {author}
  {\bibfnamefont {J.~R.}\ \bibnamefont {Rozen}}, \bibinfo {author}
  {\bibfnamefont {G.~A.}\ \bibnamefont {Keefe}}, \bibinfo {author}
  {\bibfnamefont {M.~B.}\ \bibnamefont {Rothwell}}, \bibinfo {author}
  {\bibfnamefont {M.~B.}\ \bibnamefont {Ketchen}},\ and\ \bibinfo {author}
  {\bibfnamefont {M.}~\bibnamefont {Steffen}},\ }\bibfield  {title} {\bibinfo
  {title} {Simple all-microwave entangling gate for fixed-frequency
  superconducting qubits},\ }\href
  {https://doi.org/10.1103/PhysRevLett.107.080502} {\bibfield  {journal}
  {\bibinfo  {journal} {Phys. Rev. Lett.}\ }\textbf {\bibinfo {volume} {107}},\
  \bibinfo {pages} {080502} (\bibinfo {year} {2011})}\BibitemShut {NoStop}%
\bibitem [{\citenamefont {Krantz}\ \emph {et~al.}(2019)\citenamefont {Krantz},
  \citenamefont {Kjaergaard}, \citenamefont {Yan}, \citenamefont {Orlando},
  \citenamefont {Gustavsson},\ and\ \citenamefont {Oliver}}]{Krantz2019}%
  \BibitemOpen
  \bibfield  {author} {\bibinfo {author} {\bibfnamefont {P.}~\bibnamefont
  {Krantz}}, \bibinfo {author} {\bibfnamefont {M.}~\bibnamefont {Kjaergaard}},
  \bibinfo {author} {\bibfnamefont {F.}~\bibnamefont {Yan}}, \bibinfo {author}
  {\bibfnamefont {T.~P.}\ \bibnamefont {Orlando}}, \bibinfo {author}
  {\bibfnamefont {S.}~\bibnamefont {Gustavsson}},\ and\ \bibinfo {author}
  {\bibfnamefont {W.~D.}\ \bibnamefont {Oliver}},\ }\bibfield  {title}
  {\bibinfo {title} {A quantum engineer's guide to superconducting qubits},\
  }\href {https://doi.org/10.1063/1.5089550} {\bibfield  {journal} {\bibinfo
  {journal} {Applied Physics Reviews}\ }\textbf {\bibinfo {volume} {6}},\
  \bibinfo {pages} {021318} (\bibinfo {year} {2019})},\ \Eprint
  {https://arxiv.org/abs/https://doi.org/10.1063/1.5089550}
  {https://doi.org/10.1063/1.5089550} \BibitemShut {NoStop}%
\bibitem [{\citenamefont {Suzuki}\ \emph {et~al.}(2021)\citenamefont {Suzuki},
  \citenamefont {Kawase}, \citenamefont {Masumura}, \citenamefont {Hiraga},
  \citenamefont {Nakadai}, \citenamefont {Chen}, \citenamefont {Nakanishi},
  \citenamefont {Mitarai}, \citenamefont {Imai}, \citenamefont {Tamiya} \emph
  {et~al.}}]{suzuki2021qulacs}%
  \BibitemOpen
  \bibfield  {author} {\bibinfo {author} {\bibfnamefont {Y.}~\bibnamefont
  {Suzuki}}, \bibinfo {author} {\bibfnamefont {Y.}~\bibnamefont {Kawase}},
  \bibinfo {author} {\bibfnamefont {Y.}~\bibnamefont {Masumura}}, \bibinfo
  {author} {\bibfnamefont {Y.}~\bibnamefont {Hiraga}}, \bibinfo {author}
  {\bibfnamefont {M.}~\bibnamefont {Nakadai}}, \bibinfo {author} {\bibfnamefont
  {J.}~\bibnamefont {Chen}}, \bibinfo {author} {\bibfnamefont {K.~M.}\
  \bibnamefont {Nakanishi}}, \bibinfo {author} {\bibfnamefont {K.}~\bibnamefont
  {Mitarai}}, \bibinfo {author} {\bibfnamefont {R.}~\bibnamefont {Imai}},
  \bibinfo {author} {\bibfnamefont {S.}~\bibnamefont {Tamiya}}, \emph
  {et~al.},\ }\bibfield  {title} {\bibinfo {title} {Qulacs: a fast and
  versatile quantum circuit simulator for research purpose},\ }\href@noop {}
  {\bibfield  {journal} {\bibinfo  {journal} {Quantum}\ }\textbf {\bibinfo
  {volume} {5}},\ \bibinfo {pages} {559} (\bibinfo {year} {2021})}\BibitemShut
  {NoStop}%
\bibitem [{\citenamefont {Hagberg}\ \emph {et~al.}(2008)\citenamefont
  {Hagberg}, \citenamefont {Schult},\ and\ \citenamefont {Swart}}]{networkx}%
  \BibitemOpen
  \bibfield  {author} {\bibinfo {author} {\bibfnamefont {A.~A.}\ \bibnamefont
  {Hagberg}}, \bibinfo {author} {\bibfnamefont {D.~A.}\ \bibnamefont
  {Schult}},\ and\ \bibinfo {author} {\bibfnamefont {P.~J.}\ \bibnamefont
  {Swart}},\ }\bibfield  {title} {\bibinfo {title} {{Exploring network
  structure, dynamics, and function using NetworkX}},\ }in\ \href@noop {}
  {\emph {\bibinfo {booktitle} {{In Proceedings of the 7th Python in Science
  Conference (SciPy2008)}}}}\ (\bibinfo {year} {2008})\ pp.\ \bibinfo {pages}
  {11--15}\BibitemShut {NoStop}%
\bibitem [{\citenamefont {Sheldon}\ \emph {et~al.}(2016)\citenamefont
  {Sheldon}, \citenamefont {Bishop}, \citenamefont {Magesan}, \citenamefont
  {Filipp}, \citenamefont {Chow},\ and\ \citenamefont
  {Gambetta}}]{sheldon2016characterizing}%
  \BibitemOpen
  \bibfield  {author} {\bibinfo {author} {\bibfnamefont {S.}~\bibnamefont
  {Sheldon}}, \bibinfo {author} {\bibfnamefont {L.~S.}\ \bibnamefont {Bishop}},
  \bibinfo {author} {\bibfnamefont {E.}~\bibnamefont {Magesan}}, \bibinfo
  {author} {\bibfnamefont {S.}~\bibnamefont {Filipp}}, \bibinfo {author}
  {\bibfnamefont {J.~M.}\ \bibnamefont {Chow}},\ and\ \bibinfo {author}
  {\bibfnamefont {J.~M.}\ \bibnamefont {Gambetta}},\ }\bibfield  {title}
  {\bibinfo {title} {Characterizing errors on qubit operations via iterative
  randomized benchmarking},\ }\href@noop {} {\bibfield  {journal} {\bibinfo
  {journal} {Physical Review A}\ }\textbf {\bibinfo {volume} {93}},\ \bibinfo
  {pages} {012301} (\bibinfo {year} {2016})}\BibitemShut {NoStop}%
\end{thebibliography}%
\end{document}